\pdfoutput=1
\documentclass[11pt]{article}
\usepackage{graphicx}
\usepackage{subfigure,amssymb,amsmath}
\usepackage{cite,color,xcolor,url,cancel,ulem,float}
\usepackage{bm}
\usepackage{float}
\usepackage{jheppub}

\hypersetup{linktocpage}
\newcommand{\newc}{\newcommand}

\def\Ord{\lower .7ex\hbox{$\;\stackrel{\textstyle <}{\sim}\;$}}
\def\OOrd{\lower .7ex\hbox{$\;\stackrel{\textstyle >}{\sim}\;$}}

\newc{\order}{{\cal O}}

\newc{\be}{\begin{equation}}
\newc{\ee}{\end{equation}}
\newc{\br}{\begin{eqnarray}}
\newc{\er}{\end{eqnarray}}
\newc{\ba}{\begin{array}}
\newc{\ea}{\end{array}}
\newc{\bi}{\begin{itemize}}
\newc{\ei}{\end{itemize}}
\newc{\bn}{\begin{enumerate}}
\newc{\en}{\end{enumerate}}
\newc{\bc}{\begin{center}}
\newc{\ec}{\end{center}}
\newc{\ul}{\underline}
\newc{\ra}{\rightarrow}
\newc{\lra}{\longrightarrow}
\newc{\wt}{\widetilde}
\newc{\til}{\tilde}

\newc{\wh}{\widehat}
\newc{\ti}{\times}
\newc{\Dir}{\kern -6.4pt\Big{/}}
\newc{\Dirin}{\kern -10.4pt\Big{/}\kern 4.4pt}
\newc{\DDir}{\kern -10.6pt\Big{/}}
\newc{\DGir}{\kern -6.0pt\Big{/}}
\newc{\sig}{\sigma}
\newc{\sigmalstop}{\sig_{\lstoppair}}
\newc{\Sig}{\Sigma}  
\newc{\del}{\delta}
\newc{\Del}{\Delta}
\newc{\lam}{\lambda}
\newc{\Lam}{\Lambda}
\newc{\gam}{\gamma}
\newc{\Gam}{\Gamma}
\newc{\eps}{\epsilon}
\newc{\Eps}{\Epsilon}
\newc{\kap}{\kappa}
\newc{\Kap}{\Kappa}
\newc{\modulus}[1]{\left| #1 \right|}
\newc{\eq}[1]{(\ref{eq:#1})}
\newc{\eqs}[2]{(\ref{eq:#1},\ref{eq:#2})}
\newc{\etal}{{\it et al.}\ }
\newc{\ibid}{{\it ibid}.}
\newc{\ibidem}{{\it ibidem}.}
\newc{\eg}{{\it e.g.}\ }
\newc{\ie}{{\it i.e.}\ }

\newc{\nonum}{\nonumber}
\newc{\lab}[1]{\label{eq:#1}}
\newc{\dpr}[2]{({#1}\cdot{#2})}
\newc{\lt}{\stackrel{<}}
\newc{\gt}{\stackrel{>}}
\newc{\lsimeq}{\stackrel{<}{\sim}}
\newc{\gsimeq}{\stackrel{>}{\sim}}
\def\lsim{\buildrel{\scriptscriptstyle <}\over{\scriptscriptstyle\sim}}
\def\gsim{\buildrel{\scriptscriptstyle >}\over{\scriptscriptstyle\sim}}
\def\lapp{\mathrel{\rlap{\raise.5ex\hbox{$<$}}
                    {\lower.5ex\hbox{$\sim$}}}}
\def\gapp{\mathrel{\rlap{\raise.5ex\hbox{$>$}}
                    {\lower.5ex\hbox{$\sim$}}}}
\newc{\half}{\frac{1}{2}}

\newc{\bQ}{\ol{Q}}
\newc{\dota}{\dot{\alpha }}
\newc{\dotb}{\dot{\beta }}
\newc{\dotd}{\dot{\delta }}
\newc{\nindnt}{\noindent}

\newc{\matth}{\mathsurround=0pt}
\def\ML{\ifmmode{{\mathaccent"7E M}_L}
             \else{${\mathaccent"7E M}_L$}\fi}
\def\MR{\ifmmode{{\mathaccent"7E M}_R}
             \else{${\mathaccent"7E M}_R$}\fi}

\newc{\mr}{\mathrm}

\newc{\siminf}{\mbox{$_{\sim}$ {\small {\hspace{-1.em}{$<$}}}    }}
\newc{\simsup}{\mbox{$_{\sim}$ {\small {\hspace{-1.em}{$>$}}}    }}

\newc {\Zboson}{{\mathrm Z}^{0}}
\newc{\thetaw}{\theta_W}
\newc{\mbot}{{m_b}}
\newc{\mtop}{{m_t}}
\newc{\sm}{${\cal {SM}}$}
\newc{\as}{\alpha_s}
\newc{\aem}{\alpha_{em}}

\newc{\ppbar}{\mbox{$p\ol{p}$}}
\newc{\bbbar}{\mbox{$b\ol{b}$}}
\newc{\ccbar}{\mbox{$c\ol{c}$}}
\newc{\ttbar}{\mbox{$t\ol{t}$}}
\newc{\eebar}{\mbox{$e\ol{e}$}}
\newc{\zzero}{\mbox{$Z^0$}}

\newc{\wplus}{\mbox{$W^+$}}
\newc{\wminus}{\mbox{$W^-$}}
\newc{\ellp}{\ell^+}
\newc{\ellm}{\ell^-}
\newc{\elp}{\mbox{$e^+$}}
\newc{\elm}{\mbox{$e^-$}}
\newc{\elpm}{\mbox{$e^{\pm}$}}
\newc{\qbar}     {\mbox{$\ol{q}$}}

\newc{\Ebar}{{\bar E}}
\newc{\Dbar}{{\bar D}}
\newc{\Ubar}{{\bar U}}
\newc{\susy}{{{SUSY}}}
\newc{\msusy}{{{M_{SUSY}}}}

\def\photino{\ifmmode{\mathaccent"7E \gam}\else{$\mathaccent"7E \gam$}\fi}
\def\taugluino{\ifmmode{\tau_{\mathaccent"7E g}}
             \else{$\tau_{\mathaccent"7E g}$}\fi}
\def\mphotino{\ifmmode{m_{\mathaccent"7E \gam}}
             \else{$m_{\mathaccent"7E \gam}$}\fi}
\newc{\gl}   {\mbox{$\wt{g}$}}
\newc{\mgl}  {\mbox{$m_{\gl}$}}

\def \chonep {{\wt\chi_1^+}}

\def \ch2p {{\wt\chi_2^+}}
\def \chonem {{\wt\chi_1^-}}
\def \ch2m {{\wt\chi_2^-}}

\def \chonepm{{\wt\chi_1}^{\pm}}

\def \mchonepm{m_{\chonepm}}

\newc{\dmchi}{\Delta m_{\wt\chi}}

\def \lspone{\wt\chi_1^0}
\def \mlspone{m_{\lspone}}
\def \lsptwo{\wt\chi_2^0}
\def \mlsptwo{m_{\lsptwo}}
\def \lspthree{\wt\chi_3^0}
\def \mlspthree{m_{\lspthree}}

\newc{\sele}{\wt{\mathrm e}}
\newc{\sell}{\wt{\ell}}

\def \stauone{\wt\tau_1}
\def \stautwo{\wt\tau_2}

\def \mstauone{m_{\stauone}}
\def \mstautwo{m_{\stautwo}}

\newc{\snue}     {\mbox{$ \wt{\nu_e}$}}
\newc{\smu}{\wt{\mu}}
\newc{\stau}{\wt{\tau}}
\newc {\nuL} {\wt{\nu}_L}
\newc {\nuR} {\wt{\nu}_R}
\newc {\snub} {\bar{\wt{\nu}}}
\newc {\eL} {\wt{e}_L}
\newc {\eR} {\wt{e}_R}

\def \stau{\wt\tau}

\def \sq{\wt{q}}

\newc{\msqot}  {\mbox{$m_(\sq_{1,2} )$}}
\newc{\sqbar}    {\mbox{$\bar{\wt{q}}$}}
\newc{\ssb}      {\mbox{$\squark\ol{\squark}$}}
\newc {\qL} {\wt{q}_L}
\newc {\qR} {\wt{q}_R}
\newc {\uL} {\wt{u}_L}
\newc {\uR} {\wt{u}_R}
\def \ul{\wt{u}_L}

\newc {\dL} {\wt{d}_L}
\newc {\dR} {\wt{d}_R}
\newc {\cL} {\wt{c}_L}
\newc {\cR} {\wt{c}_R}
\newc {\sL} {\wt{s}_L}
\newc {\sR} {\wt{s}_R}
\newc {\tL} {\wt{t}_L}
\newc {\tR} {\wt{t}_R}
\newc {\stb} {\ol{\wt{t}}_1}
\newc {\sbot} {\wt{b}_1}
\newc {\msbot} {m_{\sbot}}
\newc {\sbotb} {\ol{\wt{b}}_1}
\newc {\bL} {\wt{b}_L}
\newc {\bR} {\wt{b}_R}

\newc{\csquark}  {\mbox{$\wt{c}$}}
\newc{\csquarkl} {\mbox{$\wt{c}_L$}}
\newc{\mcsl}     {\mbox{$m(\csquarkl)$}}

\newc {\stopl}         {\wt{\mathrm{t}}_{\mathrm L}}
\newc {\stopr}         {\wt{\mathrm{t}}_{\mathrm R}}
\newc {\stoppair}      {\wt{\mathrm{t}}_{1}
\bar{\wt{\mathrm{t}}}_{1}}
\def \lstop{\wt{t}_{1}}

\def \lstoppair{\lstop\lstop^*}

\newc{\tsquark}  {\mbox{$\wt{t}$}}
\newc{\ttb}      {\mbox{$\tsquark\ol{\tsquark}$}}
\newc{\ttbone}   {\mbox{$\tsquark_1\ol{\tsquark}_1$}}

\newc{\mix}{\theta_{\wt t}}
\newc{\cost}{\cos{\theta_{\wt t}}}
\newc{\sint}{\sin{\theta_{\wt t}}}
\newc{\costloop}{\cos{\theta_{\wt t_{loop}}}}

\newc{\mixsbot}{\theta_{\wt b}}

\newc{\tb}{\tan\beta}
\newc{\cb}{\cot\beta}
\newc{\vev}[1]{{\left\langle #1\right\rangle}}

\newc{\mhalf}{m_{1/2}}
\newc{\mzero} {\mbox{$m_0$}}
\newc{\azero} {\mbox{$A_0$}}

\newc{\lb}{\lam}
\newc{\lbp}{\lam^{\prime}}
\newc{\lbpp}{\lam^{\prime\prime}}
\newc{\rpv}{{\not \!\! R_p}}
\newc{\rpvm}{{\not  R_p}}
\newc{\rp}{R_{p}}
\newc{\rpmssm}{{RPC MSSM}}
\newc{\rpvmssm}{{RPV MSSM}}

\newc{\sbyb}{S/$\sqrt B$}
\newc{\pelp}{\mbox{$e^+$}}
\newc{\pelm}{\mbox{$e^-$}}
\newc{\pelpm}{\mbox{$e^{\pm}$}}
\newc{\epem}{\mbox{$e^+e^-$}}
\newc{\lplm}{\mbox{$\ell^+\ell^-$}}

\def\Ecm{\ifmmode{E_{\mathrm{cm}}}\else{$E_{\mathrm{cm}}$}\fi}
\newc{\rts}{\sqrt{s}}
\newc{\rtshat}{\sqrt{\hat s}}
\newc{\gev}{\,GeV}
\newc{\mev}{~{\rm MeV}}
\newc{\tev}  {\mbox{$\;{\rm TeV}$}}
\newc{\gevc} {\mbox{$\;{\rm GeV}/c$}}
\newc{\gevcc}{\mbox{$\;{\rm GeV}/c^2$}}
\newc{\intlum}{\mbox{${ \int {\cal L} \; dt}$}}
\newc{\call}{{\cal L}}
\def \met  {\mbox{${E\!\!\!\!/_T}$}}

\newc{\ptmiss}{/ \hskip-7pt p_T}

\newc{\PT}{\mbox{$p_T$}}
\newc{\ET}{\mbox{$E_T$}}
\newc{\dedx}{\mbox{${\rm d}E/{\rm d}x$}}
\newc{\ifb}{\mbox{${\rm fb}^{-1}$}}
\newc{\ipb}{\mbox{${\rm pb}^{-1}$}}
\newc{\pb}{~{\rm pb}}
\newc{\fb}{~{\rm fb}}
\newc{\ycut}{y_{\mathrm{cut}}}
\newc{\chis}{\mbox{$\chi^{2}$}}

\def \jet(s){\emph{jet(s) }}

\newc{\mpl}{M_{\rm Pl}}
\newc{\mgut}{M_{GUT}}
\newc{\mw}{M_{W}}
\newc{\mweak}{M_{weak}}
\newc{\mz}{M_{Z}}

\newc{\OPALColl}   {OPAL Collaboration}
\newc{\ALEPHColl}  {ALEPH Collaboration}
\newc{\DELPHIColl} {DELPHI Collaboration}
\newc{\XLColl}     {L3 Collaboration}
\newc{\JADEColl}   {JADE Collaboration}
\newc{\CDFColl}    {CDF Collaboration}
\newc{\DXColl}     {D0 Collaboration}
\newc{\HXColl}     {H1 Collaboration}
\newc{\ZEUSColl}   {ZEUS Collaboration}
\newc{\LEPColl}    {LEP Collaboration}
\newc{\ATLASColl}  {ATLAS Collaboration}
\newc{\CMSColl}    {CMS Collaboration}
\newc{\UAColl}    {UA Collaboration}
\newc{\KAMLANDColl}{KamLAND Collaboration}
\newc{\IMBColl}    {IMB Collaboration}
\newc{\KAMIOColl}  {Kamiokande Collaboration}
\newc{\SKAMIOColl} {Super-Kamiokande Collaboration}
\newc{\SUDANTColl} {Soudan-2 Collaboration}
\newc{\MACROColl}  {MACRO Collaboration}
\newc{\GALLEXColl} {GALLEX Collaboration}
\newc{\GNOColl}    {GNO Collaboration}
\newc{\SAGEColl}  {SAGE Collaboration}
\newc{\SNOColl}  {SNO Collaboration}
\newc{\CHOOZColl}  {CHOOZ Collaboration}
\newc{\PDGColl}  {Particle Data Group Collaboration}

\def\issue(#1,#2,#3){{\bf #1}, #2 (#3)}
\def\iss(#1,#2,#3){{\bf #1} (#3) #2}
\def\ASTR(#1,#2,#3){Astropart.\ Phys. \issue(#1,#2,#3)}
\def\AJ(#1,#2,#3){Astrophysical.\ Jour. \issue(#1,#2,#3)}
\def\AJS(#1,#2,#3){Astrophys.\ J.\ Suppl. \issue(#1,#2,#3)}
\def\APP(#1,#2,#3){Acta.\ Phys.\ Pol. \issue(#1,#2,#3)}
\def\JCAP(#1,#2,#3){Journal\ XX. \issue(#1,#2,#3)} 
\def\SC(#1,#2,#3){Science \issue(#1,#2,#3)}
\def\PRD(#1,#2,#3){Phys.\ Rev.\ D \issue(#1,#2,#3)}
\def\PR(#1,#2,#3){Phys.\ Rev.\ \issue(#1,#2,#3)} 
\def\PRC(#1,#2,#3){Phys.\ Rev.\ C \issue(#1,#2,#3)}
\def\NPB(#1,#2,#3){Nucl.\ Phys.\ B \issue(#1,#2,#3)}
\def\NPPS(#1,#2,#3){Nucl.\ Phys.\ Proc. \ Suppl \issue(#1,#2,#3)}
\def\NJP(#1,#2,#3){New.\ J.\ Phys. \issue(#1,#2,#3)}
\def\JP(#1,#2,#3){J.\ Phys.\issue(#1,#2,#3)}
\def\PL(#1,#2,#3){Phys.\ Lett. \issue(#1,#2,#3)}
\def\ZP(#1,#2,#3){Z.\ Phys. \issue(#1,#2,#3)}
\def\ZPC(#1,#2,#3){Z.\ Phys.\ C  \issue(#1,#2,#3)}
\def\PREP(#1,#2,#3){Phys.\ Rep. \issue(#1,#2,#3)}
\def\PRL(#1,#2,#3){Phys.\ Rev.\ Lett. \issue(#1,#2,#3)}
\def\MPL(#1,#2,#3){Mod.\ Phys.\ Lett. \issue(#1,#2,#3)}
\def\RMP(#1,#2,#3){Rev.\ Mod.\ Phys. \issue(#1,#2,#3)}
\def\SJNP(#1,#2,#3){Sov.\ J.\ Nucl.\ Phys. \issue(#1,#2,#3)}
\def\CPC(#1,#2,#3){Comp.\ Phys.\ Comm. \issue(#1,#2,#3)}
\def\IJMPA(#1,#2,#3){Int.\ J.\ Mod. \ Phys.\ A \issue(#1,#2,#3)}
\def\MPLA(#1,#2,#3){Mod.\ Phys.\ Lett.\ A \issue(#1,#2,#3)}
\def\PTP(#1,#2,#3){Prog.\ Theor.\ Phys. \issue(#1,#2,#3)}
\def\RMP(#1,#2,#3){Rev.\ Mod.\ Phys. \issue(#1,#2,#3)}
\def\NIMA(#1,#2,#3){Nucl.\ Instrum.\ Methods \ A \issue(#1,#2,#3)}
\def\EPJC(#1,#2,#3){Eur.\ Phys.\ J.\ C \issue(#1,#2,#3)}
\def\RPP (#1,#2,#3){Rept.\ Prog.\ Phys. \issue(#1,#2,#3)}
\def\PPNP(#1,#2,#3){ Prog.\ Part.\ Nucl.\ Phys. \issue(#1,#2,#3)}
\newc{\PRDR}[3]{{Phys. Rev. D} {\bf #1}, Rapid  Communications, #2 (#3)}

\def\PLB(#1,#2,#3){Phys.\ Lett.\ B  \iss(#1,#2,#3)}
\def\JHEP(#1,#2,#3){JHEP \iss(#1,#2,#3)}

\def\gmin2{(g-2)_\mu}

\catcode`\@=11 

\def\vev#1{\left\langle #1\right\rangle}
\def\lsim{\mathrel{\mathpalette\@versim<}}
\def\gsim{\mathrel{\mathpalette\@versim>}}
\def\@versim#1#2{\vcenter{\offinterlineskip
    \ialign{$\m@th#1\hfil##\hfil$\crcr#2\crcr\sim\crcr } }}
\def\etal{{\em et. al.}}

\def\r2{\sqrt 2}
\def\beq{\begin{equation}}
\def\eeq{\end{equation}}
\def\beqn{\begin{eqnarray}}
\def\eeqn{\end{eqnarray}}
\def\sinW2{\sin^2\theta_W}

\def\mz2{M_{z}^2}
\def\c2b{\cos 2\beta}

\def\m#1{{\tilde m}_#1}

\def\mw#1{{\tilde m}_{\omega #1}}

\def\mz{M_Z}
\def\m0{m_0}
\def\mhalf{m_{\frac{1}{2}}}

\def\cb{\cos\beta}
\def\sb{\sin\beta}

\def\sec2w{sec^2\theta_W}

\def\gmin2{(g-2)_\mu}

\catcode`\@=11 

\def\vev#1{\left\langle #1\right\rangle}
\def\lsim{\mathrel{\mathpalette\@versim<}}
\def\gsim{\mathrel{\mathpalette\@versim>}}
\def\@versim#1#2{\vcenter{\offinterlineskip
    \ialign{$\m@th#1\hfil##\hfil$\crcr#2\crcr\sim\crcr } }}
\def\etal{{\em et. al.}}

\def\tb{\tilde b}

\def\tL{\tilde L}

\def \chonep{{\wt\chi_1}^{+}}
\def \chonem{{\wt\chi_1^-}}
\def \chonep2{{\wt\chi_2^+}}
\def \chonem2{{\wt\chi_2^-}}

\def \chonepm{{\wt\chi_1}^{\pm}}

\def \mchonepm{m_{\chonepm}}

\def \lstop{\wt{t}_{1}}

\def \lspone{\wt\chi_1^0}
\def \mlspone{m_{\lspone}}
\def \lsptwo{\wt\chi_2^0}
\def \mlsptwo{m_{\lsptwo}}
\def \lspthree{\wt\chi_3^0}
\def \mlspthree{m_{\lspthree}}

\def\PL{Phys. Lett.}
\def\PRL{Phys. Rev. Lett.}

\def\PR{Phys. Rev.}

\def \lsptwo{\wt\chi_2^0}
\def \lspone{\wt\chi_1^0}
\def \chonem {{\wt\chi_1^\pm}}
\def \chargino1 {{\wt\chi_1^\pm}}
\def \chargino2 {{\wt\chi_2^\pm}}
\def \lstop{\wt{t}_{1}}
\def \ch2m {{\wt\chi_2^-}}

\def \chonep {{\wt\chi_1^+}}

\def\mygraph#1#2{ \subfigure[]{
   \label{#1}
   \hspace*{-0.6in}
   \begin{minipage}[b]{0.5\textwidth}
   \centering
   \hspace*{4ex}
   \includegraphics[width=1.2\textwidth]{#2}
   \vspace*{-4ex}
   \end{minipage}}
   \vspace*{-1ex}}

\def\mygraphthree#1#2{ \subfigure[]{
   \label{#1}
   \hspace*{-0.6in}
   \begin{minipage}[b]{0.5\textwidth}
   \centering
   \hspace*{4ex}
   \includegraphics[width=1.0\textwidth]{#2}
   \vspace*{-4ex}
   \end{minipage}}
   \vspace*{-1ex}}

\title{
How light a higgsino or a wino dark matter can become in a compressed  scenario of MSSM
}
\author[a]{Manimala Chakraborti,}
\author[b]{Utpal Chattopadhyay,}
\author[c]{Sujoy Poddar}
\affiliation[a]{
Bethe Center for Theoretical Physics \& Physikalisches Institut der Universit\"at Bonn, 
Nu{\ss}allee 12, 53115 Bonn, Germany
}

\affiliation[b]{
Department of Theoretical Physics, Indian Association
for the Cultivation of Science,
2A \& B Raja S.C. Mullick Road, Jadavpur,
Kolkata 700 032, India
}

\affiliation[c]{
Department of Physics, Netaji Nagar Day College, 170/436, N.S.C. Bose Road, 
Kolkata - 700092, India
}

\emailAdd{mani.chakraborti@gmail.com}
\emailAdd{tpuc@iacs.res.in}
\emailAdd{sujoy.phy@gmail.com}

\abstract{
 Higgsinos and Wino have strong motivations for being 
  Dark Matter (DM) candidates in supersymmetry, but their annihilation cross sections
  are quite large.
    For thermal generation and a single component DM setup 
the higgsinos or wino may have masses of around 1 or 2-3 TeV respectively.
  For such DM candidates, 
  a small amount of slepton coannihilation may decrease the effective
  DM annihilation cross section.
  This, in turn reduces the lower limit of the relic density satisfied 
  DM mass by more than 50\%. 
  Almost a similar degree of reduction of the same limit is also seen for
  squark coannihilations. However, on the contrary, for
  near degeneracy of squarks and higgsino
  DM, near its generic upper limit, the associated coannihilations may decrease 
  the relic density, thus extending the upper limit
  towards higher DM masses. 
  We also compute the direct and indirect detection signals.
  Here, because of the quasi-mass degeneracy of the squarks and the
    LSP, we
  come across a situation where squark exchange diagrams
  may contribute significantly or more strongly than the
  Higgs exchange contributions in the spin-independent
  direct detection cross section of DM.
  For the higgsino-DM scenario,
  we observe that a DM mass of 600 GeV to be consistent with
  WMAP/PLANCK and LUX data for sfermion coannihilations.
  The LUX data itself excludes the region of 450 to 600 GeV,
  by a half order of magnitude of the cross-section, well below
  the associated uncertainty.
  The similar combined lower limit for
  a wino DM is about 1.1 TeV.
  There is hardly any collider bound from the LHC for squarks and
    sleptons in such a compressed scenario where sfermion masses are close
    to the mass of a higgsino/wino LSP.
}

\begin{document}
\maketitle
\section {Introduction}
\label{intro}
One of the most interesting features of low energy supersymmetry
(SUSY)\cite{susyreviews1,susyreviews2,susybooks} 
is that it can provide with a viable
candidate for dark matter (DM)\cite{Kamionkowski, Silk}.  The lightest supersymmetric particle
(LSP), typically the lightest neutralino, 
in R-parity conserving scenarios of
the Minimal Supersymmetric Standard Model
(MSSM)\cite{susyreviews1,susyreviews2,susybooks} is a strong candidate for
DM. In MSSM, the neutralinos are composed of gauginos
(bino and wino) and higgsinos. 
Whether the LSP in a particular MSSM scenario is able to satisfy the right relic abundance,
depends crucially on its gaugino-higgsino compositions. For example, in the minimal supergravity model (mSUGRA)\cite{susybooks,msugra_orig} that  
assumes SUSY breaking at a high scale, 
the LSP is bino-like for most of the parameter space. Bino being a gauge singlet, its
annihilation occurs mainly through t-channel exchange of sfermions, giving rise to the {\it bulk annihilation} region\cite{susyreviews2}.  With the discovery of a 125 GeV Higgs 
boson at the Large Hadron Collider (LHC) of CERN\cite{higgs}, the 
sfermion masses are pushed towards higher values, particularly for the 
universal models like mSUGRA.  This makes bulk annihilation rather an
inefficient mechanism 
for obtaining the right relic density\cite{dmsugra_recent}. 
It is however possible to have a bino-LSP giving the correct abundance of  
DM in some specific regions of parameter space of mSUGRA, like the stau coannihilation, 
funnel (resonant Higgs annihilation) or hyperbolic branch (HB)\cite{HB,HBnew}/
focus point (FP)\cite{FP} regions. 
While most of the former two regions of parameter space are ruled out by the Higgs data\cite{higgs}, 
a large part of the HB/FP region, which corresponds to a significant 
degree of bino-higgsino mixing,
is disfavored by DM direct detection experiments\cite{dd_msugra}.
There is also a higgsino LSP region in 
mSUGRA but this comes with a baggage of a very large 
gluino mass and a heavy SUSY spectra in general. A SUSY scenario where
the spectra may not necessarily be so heavy while the LSP is still highly 
higgsino dominated in nature may be realized in a
few nonuniversal gaugino mass models 
\cite{etcEllis:1985jn,NathMixedRep,
Chattopadhyay:2001mj,nugminter,ucdphiggsino,nugmvarious,
Chattopadhyay:2009fr,Guchait:2011andothers,Chattopadhyay:2005mv,
mc-uc-sr-dp-2014}, nonuniversal Higgs scalar
models\cite{NUHMhiggsino} or models with non-holomorphic soft terms\cite{NHSSMhiggsino}.

Apart from bino, one may have the right relic 
abundance in MSSM if the LSP is a higgsino,  
wino or a well-managed admixture of bino, wino and
higgsinos\cite{arkani-hamed}. 
For the higgsino-like LSP scenario, the lightest neutralino 
($\lspone$) mass is close to the value of the higgsino mixing 
parameter $\mu$. In this case the lightest 
chargino $\chonepm$ and the second lightest neutralino 
$\lsptwo$ are almost mass degenerate with 
$\lspone$ with their masses are close to $\mu$.  For a thermally generated
single component dark matter,
it has been {\it typically accepted} that 
the LSP mass for obtaining the right DM abundance is around 
1 TeV for a higgsino-like LSP\cite{HBnew,dd_msugra}. 
Below this limit the annihilation and coannihilations among $\lspone$,
$\lsptwo$ and $\chonepm$ are too strong causing the DM relic density to become underabundant.

On the other hand, a wino-like LSP 
may be possible to realize if $M_2 < (M_1,\mu)$ where $M_2$ and $M_1$ 
are the $SU(2)_L$ and $U(1)_Y$ gaugino mass 
parameters respectively.
A wino may be the candidate for LSP   
in many theoretically well-motivated models like the
Anomaly Mediated Supersymmetry Breaking (AMSB)\cite{AMSBorig}.
For a wino LSP, $\mlspone$ and $\mchonepm$ lie very
close to $M_2$ allowing strong coannihilations. 
A thermally generated wino dark matter is underabundant up to $\sim$~ 2 TeV. 
It satisfies the relic density data for a wino mass of 
2-3 TeV\footnote{The spread of results depends on purity of
  wino, the extent of decoupling of the squark masses as well as
  the higgsino mass parameter $\mu$,
  inclusion of non-perturbative effects like Sommerfeld corrections
  etc. We will
come back to it in Sec.\ref{subsec_wino}.}\cite{uc-dd-pk-dp-wino,matsumoto_wino,Beneke:2016ync,Cirelli:2007xd,generalwinodm}.

Focusing on the collider information as received from the LHC 
we note that the SUSY searches at the LHC in generic MSSM spectra i.e. in 
an uncompressed sparticle mass scenario,  
have imposed stringent bounds on the
masses of strong sector sparticles\cite{atlasall,cmsall}.  The strong sector 
scalar masses are increasingly pushed above a TeV regime that is even 
superseded by the gluino mass limits.
On the other hand, the direct mass
bounds on the electroweak (EW) sector sparticles from the LHC searches are 
rather mild\cite{lhcew1,lhcew2}.
In the context of mass limits of sparticles, we must 
however remember that the LHC searches
are restricted to the so-called ``simplified models'' that are
characterized by certain assumptions on sparticle masses and compositions of
the EW sector gauginos (electroweakinos). 
The searches in the 3$l + \met$ channel\cite{atlas3l} for example 
consider $\lspone$ to be purely a bino and $\chonepm / \lsptwo$
to be completely wino dominated.
Imposing basic 
constraints like the Higgs mass, dark matter relic density
and muon $g-2$ there have been studies that effectively probed the SUSY 
parameter space for the above types of 
electroweakinos\cite{ourwork1,otherewrefs,otherewrefs2}. 
However, it turns out that the collider limits get significantly 
degraded once we start varying the composition of the 
electroweakinos.  This may be seen in Ref.\cite{ourwork2}  
where the authors considered  
$\chonepm$ to be higgsino dominated or a mixture of a wino and higgsinos in a 
bino dominated LSP ($\lspone$) situation. 
Similar to the above, changed composition of the LSP itself 
may significantly alter the collider limits. 
For example, the trilepton search limits are hardly of any importance 
in a higgsino dominated LSP scenario where
$\lspone$, $\lsptwo$ and $\chonepm$ are almost mass degenerate.  
This is simply because the resulting leptons come out to be very soft. 
For collider studies of benchmark points that 
satisfy the observed relic density
range, one may however use monojet + $\met$ 
analysis\cite{lhcdmsearch}. However, the bounds are seen to
be very weak \cite{barducci}. Apart from all the above, collider bounds 
of sfermions including also squarks get severely diluted if one considers a 
compressed scenario of sparticle masses where the LSP is higgsino/wino 
dominated in its composition with its mass close to that of sleptons 
and/or squarks as appropriate in a LSP-sfermion coannihilation study.  

In this analysis we use a
compressed SUSY scenario in a 
phenomenological MSSM (pMSSM)\cite{Djouadi:1998di} framework so as to 
explore how light the higgsinos and wino can become while having relic 
density values within the phenomenologically accepted range. We will  
consider appropriate coannihilations of the LSP separately with
sleptons and squarks or both of them together.   

We will now briefly outline the commonly explored coannihilation scenarios. 
To determine the relic density including coannihilations
one computes a thermally averaged {\it effective} annihilation
cross-section $<\sigma_{eff} v>$ for the LSP\cite{Griest:1990kh}.
$<\sigma_{eff} v>$ is obtained from self-annihilation and various
  coannihilation cross-sections that are weighted
  by factors exponentially suppressed by relative 
  mass differences between the DM and the coannihilating partners.
The DM relic density is inversely proportional 
to $<\sigma_{eff} v>$.
In the context of mSUGRA, excluding a few specific regions 
of parameter space, the LSP is generally bino
($\tilde B$)-dominated
in its composition\cite{susybooks}. Except the t-channel slepton exchange that requires very
light sleptons or the s-channel Higgs modes, a bino typically undergoes a 
tiny amount of self-annihilation that leads to overabundance in most of the
mSUGRA parameter space. The situation changes when 
bino-slepton or in particular bino-stau coannihilation comes 
into the picture. Here the non-relativistic threshold S-wave 
coannihilation cross-sections such as 
$\sigma_{{\tilde B}-l_R}$ and $\sigma_{l_R-l_R^*}$, $l_R$ being the right handed
stau, are not suppressed by 
fermionic mass factors.  The above coannihilation cross-sections 
are much larger than the self-annihilation cross-section 
of a highly bino-dominated LSP\cite{Ellis:1999mm}.  
Thus, in mSUGRA, the coannihilation of 
$\lspone$ with ${\tilde \tau}_1 $
is able to reduce the relic density to fall within the 
WMAP/PLANCK\cite{wmap,planck} specified range. 
A detailed analysis was made in Ref.\cite{Edsjo:2003us} 
where all possible kinds of 
coannihilations were considered in a bino-dominated LSP scenario in an
mSUGRA setup. 
However, we must remember that masses of sparticles in mSUGRA are 
correlated that causes mostly LSP-stau coannihilations to be relevant. 
Thus, a significant amount of change in the DM  
relic density via coannihilations,
leads to an acceptable value, but this is possible 
only in a limited zone of parameter space.

Models with essentially 
unconstrained sparticle masses such as pMSSM when considered 
in a compressed scenario are able to probe 
the true potential of coannihilations. Here, 
the LSP may find several coannihilating 
partners almost mass-degenerate with itself that would overcome 
the associated exponential suppression encountered in computing the effective
annihilation cross section. In this analysis, we will focus
on higgsino and wino dominated LSPs that undergo coannihilations 
with sfermions, both sleptons and squarks, separately or together.
We will see that there may be rather uncommon occurrences  
when coannihilations may potentially cause a {\it decrease} in the 
effective annihilation cross-sections, thereby causing 
an increase in the DM relic density. 
This was discussed in Refs.\cite{edsjo-gondolo,profumo} where the latter 
reference named the coannihilating species as {\it parasite degree of 
freedom} in which the authors investigated the role of sleptons  
coannihilating with the LSP. In this work, we will systematically 
analyze the effects of coannihilations with all the sfermions in a compressed scenario  
and probe the mass reach of the LSP as a higgsino/wino in relation to 
the latest phenomenological constraints involving dark matter, Higgs mass and
the relevant bounds from collider data.     

We emphasize that the effect of considering a coannihilating particle,  
in particular whether
it would cause a decrease or increase of the relic density  
depends on several factors\cite{Griest:1990kh,edsjo-gondolo},
namely, i) the annihilation cross section 
$\sigma_{\chi_i^\prime-\chi_0}$ of the  
coannihilating particle $\chi_i^\prime$ with the 
LSP ($\chi_0$), ii) the cross section  
$\sigma_{\chi_i^\prime-\chi_i^\prime}$ for the coannihilating particle annihilating 
with itself, iii) $\sigma_{\chi_i^\prime-\chi_j^\prime}$, where $i$ and $j$ refer to 
different species of coannihilating particles, iv) the relative 
mass gap between 
the sparticles namely, $\delta_i=\frac{m_{\chi_i^\prime}-m_{\chi_0}}{m_{\chi_0}}$ 
or $\delta_{ij}=\frac{m_{\chi_i^\prime}-m_{\chi_j}}{m_{\chi_i}}$, 
thereby on the extent of 
exponential suppression, v) 
appropriate weight factors arising out of the degrees of freedom of the 
associated particles undergoing coannihilations.  
   
We must remember that while a bino does not have any 
gauge charge, a higgsino (wino) is associated with 
isospin $\frac{1}{2}$ ($1$) . This translates into 
a larger internal degrees of freedom, namely 8 for a higgsino and 6 for a 
wino type of LSP considering its Majorana nature. Expectedly, a large 
number of internal degrees of freedom contributes more 
toward the self-annihilation cross section of the LSP.
In addition to the above, one must also consider 
large coannihilations involving candidates like $\chonepm$, $\lsptwo$ 
for a higgsino type of LSP and $\chonepm$ for a wino type of the same.   
All the above lead to a substantially large effective annihilation cross-section 
for the above two types of LSP. Among the coannihilating 
sfermions, the left handed ones have larger internal 
degrees of freedom. This further gets multiplied by the
  color degrees of freedom for squarks.  In computing the effective
annihilation rates 
out of individual cross-sections, one notes that the 
associated weight factors for different coannihilating species 
  play very important roles in either decreasing or increasing the
  total rate itself.
As mentioned before, unlike a bino LSP, a higgsino or a wino LSP is
intrinsically associated with a larger amount of self-annihilation as well as
LSP-electroweakino coannihilations.
We will refer this as a generic higgsino or wino DM scenario.
It turns out that in 
a compressed sfermion scenario all the
appropriate degrees of freedom of the coannihilating sfermions
may contribute to the averaging  
process toward $<\sigma_{eff}v>$ in such a way that the latter becomes 
smaller than the case of having no sfermion coannihilations over most
of the parameter space satisfying the DM relic density limits. 
Thus, for a given LSP mass the relic density increases. 
This on the other hand is synonymous with a decreased lower limit of the  
mass of LSP satisfying the WMAP/PLANCK data. However, apart from
  the typical trend mentioned above, we will come across a parameter
  region corresponding to a higgsino as LSP where the outcome due to
  squark-squark coannihilations may become dominant over
  the electroweakino part of the LSP depletion cross section.

  We will also study the direct and indirect
detection prospects for the types of 
DM considered in this analysis.  The LUX\cite{luxnew} experiment
puts strong bounds on spin-independent (SI) DM direct detection 
cross-sections. 
The $\lspone-$nucleon scattering cross-section that LUX relies on is 
enhanced for sufficient 
gaugino-higgsino mixing\cite{hisano_nojiri}. However, in our scenario
the cross section is supposed to be small for LSP being so pure, either a higgsino 
or a wino.  On the other hand, the DM indirect detection 
experiments\cite{icecube1,icecube2} 
look for signals coming from stable final state particles of DM
annihilation processes in the solar or galactic cores.  
Since the sfermion-coannihilations make the smaller DM mass zones 
to become valid in relation to the relic density data, it is important to
find whether 
the indirect detection rates can also be large for much smaller
values of higgsino or wino masses satisfying the DM relic density limits.

The paper is organized as follows. In Section~\ref{sec2} we briefly 
discuss the effect of sfermion coannihilations in the calculation of the effective 
cross-section.
In Section \ref{sec_relicdensity}  we present the relic density computation results for 
higgsinos and wino types of LSP  by considering slepton and squark coannihilations
separately or together.
We will start the section by discussing the roles of the relevant
electroweakino mass differences that potentially affect the higgsino and wino relic density
results. 
Constraints coming from the direct and 
indirect DM detection experiments on our results
are discussed in Section \ref{detectionDM}. 
Finally, we conclude in Section \ref{conclusion}.   

\section {Sfermion Coannihilations with Higgsino and Wino Types of LSP}
\label{sec2}
Let us consider the evolution of a class of particles $\chi_i$, $i=1, ...N$, 
in the Early Universe. We assume the particles are different from
SM candidates by assuming an R-parity conserved scenario   
of SUSY.
The Boltzmann equation governing the 
number density $n_i$ of the i-th kind of particle at any instant of 
time $t$ is given by\cite{edsjo-gondolo},
\begin{eqnarray} \label{boltzmann1}
  \frac{dn_{i}}{dt}
  &=&
  -3 H n_{i}
  - \sum_{j=1}^N \langle \sigma_{ij} v_{ij} \rangle
    \left( n_{i} n_{j} - n_{i}^{\rm{eq}} n_{j}^{\rm{eq}} \right)
  \nonumber \\
  & &
  - \sum_{j\ne i}
  \big[ \langle \sigma'_{Xij} v_{ij} \rangle
        \left( n_i n_X - n_{i}^{\rm{eq}} n_{X}^{\rm{eq}} \right)
      - \langle \sigma'_{Xji} v_{ij} \rangle
        \left( n_j n_X - n_{j}^{\rm{eq}} n_{X}^{\rm{eq}} \right)
  \big]
  \nonumber \\
  & &
  - \sum_{j\ne i}
  \big[ \Gamma_{ij}
        \left( n_i - n_i^{\rm{eq}} \right)
      - \Gamma_{ji}
        \left( n_j - n_j^{\rm{eq}} \right)
  \big],
\end{eqnarray}
where the first term is due to the expansion of the universe and 
$H$ is the Hubble parameter\cite{Silk}. The second term arises because of 
coannihilations between $i$-th and $j$-th sparticles leading to SM
particles in the final state i.e. for processes like 
$\chi_i \chi_j \rightarrow X$.
The total cross-section for scattering off 
the cosmic thermal background, 
$\chi_i X \rightarrow \chi_j Y$ is given by $\sigma'_{Xij}$, where  
X and Y are SM type of particles.
The last term describes the decay of $\chi_i$ and  
$\Gamma_{ij}$ refers to the total decay width for the processes
$\chi_i \rightarrow \chi_j X$.
Since in an R-parity conserving scenario all the 
existing sparticles will eventually decay into the 
LSP, its number density is given as $n = \sum_{i=1}^N n_i$. 
Now, an assumption for the distribution of $\lspone$ 
maintaining its
thermal equilibrium value i.e.
$\frac{n_{i}}{n} \simeq \frac{n_{i}^{\rm{eq}}}{n^{\rm{eq}}}$, Eq. \ref{boltzmann1} leads to,
\begin{equation} \label{boltzmann3}
  \frac{dn}{dt} =
  -3Hn - \langle \sigma_{\rm{eff}} v \rangle
  \left( n^2 - n_{\rm{eq}}^2 \right),
\end{equation}
where, 
\begin{equation} \label{sigmaveffdef}
  \langle \sigma_{\rm{eff}} v \rangle = \frac{\sum_{ij} \langle
  \sigma_{ij}v_{ij} \rangle {n_{i}^{\rm{eq}}}{n_{j}^{\rm{eq}}}}
  {{n^{\rm{eq}}}^2}.
\end{equation}
In the non-relativistic approximation, one has,
\begin{equation}
\frac{n_i^{eq}}{n^{eq}} = \frac{g_i \exp(-x\delta_i)(1+\delta_i)^{3/2}}{g_{eff}},
\label{numbdensityrelation}
\end{equation}
where $g_i$ is the number of internal degrees of freedom of the $i$-th particle,
$\delta_i = \frac{m_i - m_1}{m_1}$ (for $i>1$), $x=\frac {m_1} {T}$, $m_1$ being the mass of
the LSP and 
$g_{eff} = \sum_{i=1}^{N}{g_i \exp(-x\delta_i)(1+\delta_i)^{3/2}}$. 

  In this analysis with slepton and squark coannihilations each
$\delta_i$ corresponding to a sfermion type $i$ is allowed to vary up to a
chosen limit $\delta_{max}$. Thus, each sfermion mass $m_{{\tilde f}_i}$ will
have an upper limit of $\mlspone(1+\delta_{max})$. $\delta_{max}$ is chosen as
20\% keeping in mind
the exponential suppression
within Eq.\ref{numbdensityrelation}.
Considering a slepton coannihilation scenario in pMSSM, as an example,
we note that the first two
  generations of sleptons do not differ much in their mass values among
  themselves and these will be close to $M_{\tilde l}~(=M_{{\tilde l}_R})$,
  the pMSSM common slepton
mass parameter\footnote{Only small differences come
  from the D-term contributions of the left and the right sleptons as well as
  sneutrinos.}. First, let us consider only the sleptons
of the first two generations to undergo coannihilations.
For a given $\mlspone$ and $\delta_{max}$, the sleptons
will have a maximum mass value of $\mlspone(1+\delta_{max})$ and this
will be close to the maximum value of $M_{\tilde l}$.
For a given LSP mass, calling
the average of all the associated $\delta_i$ values
as $\delta$, one finds that the highest
possible $\delta$ becomes
close to $\delta_{max}$, the chosen degree of maximum relative mass deviation,
irrespective of $\tan\beta$.  Next, we allow
the third generation of sleptons to have mass values
in the coannihilation zone.   Consequent of the L-R mixing
effect (which is more prominent
for a larger $\tan\beta$), $m_{{\tilde \tau}_1}$ and
  $m_{{\tilde \tau}_2}$ are largely separated among themselves. Only the
heavier stau will have its maximum mass value near
$\mlspone(1+\delta_{max})$ and all other sleptons will have much smaller
masses. Thus, the reach of $M_{\tilde l}$ becomes smaller and so is the
average value of all the three generations of slepton masses. Hence,
the average relative deviation $\delta$ will have its maximum value
significantly smaller than $\delta_{max}$, an effect that would increase
with $\tan\beta$.
Additionally, when LSP mass is in the smaller zone meaning a
stronger degree of electroweakino coannihilations, the requirement
of slepton coannihilations increases so as the satisfy the relic density
limits. Thus, the slepton masses are needed to stay within the close vicinity
of $\mlspone$. Consequently, $\delta$ as well as its range of variation both
become smaller for a given mass of the LSP. Here, the aforesaid
range becomes smaller because of the stau L-R mixing since
coannihilation effects
of both the staus are quite required in the process. In other words, the
stau masses cannot be too far away from $\mlspone$, or the associated
$\delta_i$ values can not be large, thus avoiding an
exponential suppression. It follows that the
minimum mass of the LSP satisfying Eq.\ref{planckdata} becomes
larger for a larger $\tan\beta$.  Similarly, squark coannihilations,
in principle, will also show qualitatively identical behaviour\footnote{We will however point out a difference for the higgsino LSP case
  in the heavier limit of the LSP.} based on the availability of the
all the generation of squarks for coannihilations.
However, we will consider
only the first two generations of squarks in this study
keeping the top-squarks in a decoupled zone because of a very large L-R mixing,
particularly arising from the requirement of satisfying the Higgs mass
data.

  In regard to a book-keeping of the internal
  degrees of freedom ({\it d.o.f.}) we note that for a given generation of
  the right and left
handed sleptons like ${\tilde \tau}_{R,L}$ there are 2 internal
{\it d.o.f.} available for each R and L while the sneutrino
${\tilde \nu}_\tau$ along with its anti-particle would have one internal
{\it d.o.f.} each. Thus, for a compressed slepton spectra undergoing
coannihilations with the LSP the total number of internal {\it d.o.f.}
for all the three generations of sleptons would be 18.
For the squark coannihilations with only two generations
  are considered in the analysis, the resulting internal
  {\it d.o.f.} amounts to 48 after accounting for the color {\it d.o.f.}.

\noindent
\section {Results for Relic Density}
\label{sec_relicdensity}

In spite of the fact that the
PLANCK\cite{planck} data for the DM relic density has a very small
uncertainty ($\Omega_{CDM} h^2=0.1199 \pm 0.0022$), we
note that there is about a 
10\% level of theoretical uncertainty
in computing the SUSY DM relic density\cite{klasen}. This is approximately
six times the
observational uncertainty, as concluded in Ref.\cite{klasen}.
It was shown that higher order SUSY-QCD corrections may cause a significant
degree of shift of the relic density in some scenarios and
the uncertainty arising out of renormalization scheme and scale variations
can be quite significant.  Several recent
analyses used such a degree of theoretical error or even more 
(see for example Refs.\cite{Bertone:2015tza,Badziak:2017uto}).
Thus, we will use $\Omega_{\lspone} h^2=0.12 \pm 0.012$ 
that leads to the following bounds. 
\begin{equation}
  0.108 < \Omega_{\lspone} h^2 < 0.132.
  \label{planckdata}
\end{equation}

We use the code {\tt SuSpect} (version 2.43)\cite{suspect} for spectrum generation.
For the calculation of DM relic density and direct and indirect detection
observables we use the code {\tt micrOMEGAs} version 3.2\cite{micromegas}.
We have also verified agreement with the code
{\tt DarkSUSY}\cite{darksusy} by choosing various representative points over the
parameter space.

Throughout the analysis we impose the Higgs mass range of
  122 to 128 GeV 
considering the uncertainty in computing SUSY Higgs
mass $m_h$ with radiative corrections\cite{higgsuncertainty3GeV}.

\subsection{Quasi degeneracy of electroweakino masses}
\label{massdiff}
Since the electroweakino coannihilations play a dominant 
  role in computing the DM relic density both for higgsino and wino types
  of LSPs, it is important to
  discuss briefly the role of the appropriate electroweakino mass
  differences\cite{Manyrefsmassdiff1,Giudice:1995np,
    Cheng:1998hc,Bomark:2013nya,Gherghetta:1999sw,Feng:1999fu}.
  In regard to a higgsino type of LSP,  
  both the mass differences $\mchonepm-\mlspone$ 
  and $\mlsptwo-\mlspone$ are important. Typically the latter is about
  double the former at the tree level\cite{Giudice:1995np}.
  For a higgsino type of LSP,
  with $M_2>\mu,M_W$, an expansion in $1/{M_2}$ leads to the
 following tree level mass difference\cite{Bomark:2013nya}.  
\begin{equation}
\mchonepm-\mlspone = \left[\frac{M_2}{M_1}\tan^2\theta_W+1+{\rm sgn}\,\mu\left(\frac{M_2}{M_1}\tan^2\theta_W-1\right)\sin{2\beta}\right]\frac{M_W^2}{2M_2}+{\mathcal O}\left(\frac{1}{M_2^2}\right).\label{eq:dmhiggsino}
\end{equation}

\noindent
In the wino limit of the LSP, one has $M_2<|\mu|,|M_1|$
causing the difference $\mchonepm-\mlspone$
to become small.  
An expansion in $1/\mu$
  leads to the following tree level relationship\cite{Bomark:2013nya}.
\begin{eqnarray}
\mchonepm-\mlspone &=& \frac{M_W^2}{\mu^2}\frac{M_W^2}{M_1-M_2}\tan^2\theta_W\sin^2{2\beta}+2\frac{M_W^4M_2\sin{2\beta}}{(M_1-M_2)\mu^3}\tan^2\theta_W\nonumber\\
&&+\frac{M_W^6\sin^3{2\beta}}{(M_1-M_2)^2\mu^3}\tan^2\theta_W(\tan^2\theta_W-1)+{\mathcal O}\left(\frac{1}{\mu^4}\right).
\label{eq:dmwinotree}
\end{eqnarray}
Additional suppression comes for large $\tan\beta$ since
$\sin^2{(2\beta)} \sim 4/{\tan^2\beta}$. Thus, 
the terms up to the order $1/\mu^3$ are suppressed indicating the
lowest contributing order to be $1/\mu^4$ which is given as follows\cite{Gherghetta:1999sw}. 
\begin{eqnarray}
  \mchonepm-\mlspone=\frac{M_2 M_W^4}{2\mu^4} \left(1+
  \frac{2M_2 \tan^2\theta_W}{M_1-M_2} \right) +
       {\mathcal O}\left(\frac{1}
       {\mu^6}\right), \quad \quad {\rm for} \tan\beta \rightarrow \infty.
\end{eqnarray}  
One finds that
the above mass splitting is small even for moderate values of $\mu$. 
Thus, the radiative corrections for the two electroweakino masses or
rather that of their difference    
become important\cite{Eberl2007-8,CharginoLoopCorrEtcOrig}. The dominant
corrections to the masses come from top-stop and
$\gamma(Z)$-higgsino loops\cite{Eberl2007-8}. On the other hand, 
the renormalization of the mass difference is controlled by the gauge
boson loops as pointed out in Refs.\cite{Cheng:1998hc,Feng:1999fu}. However,
in our analysis we find a non-negligible reduction in the mass gap
when $\mu$ and/or squark
masses are taken to be very large ($\sim 10$~TeV) and we 
agree with the conclusion of Ref.\cite{Giudice:1995np} in this regard. 
In this analysis, since we are looking for slepton and squark
coannihilations with the LSP while trying to probe the lower mass limit
of the latter, we undertake a minimalistic approach
of considering not too much different mass values for the scalars among
themselves. We also limit $\mu$ so as to have only an adequate
degree of wino purity of the LSP while trying to respect
naturalness\cite{HB,HBnew,FP,old_naturalness} as
far as possible. Hence, we will prefer not to take very
large values for the scalar masses or $\mu$.  

\subsection{Higgsino dominated LSP}
\label{subsec_higgsino}
A higgsino dominated LSP with mass close to $\mu$ 
can be made out of the choice $M_2>M_1>\mu$.
We will quantify the degree of higgsino
content of the LSP via $Z_h$ defined as   
$Z_h = (N_{13}^2 + N_{14}^2)$, where $N_{ij}, i,j = 1,2,3,4$ are
the elements of the neutralino-mass diagonalizing matrix\cite{susybooks}.
In particular, we have used 
the following choice: 
$M_1 = 2 \mu$ and $M_2 =2.4 \mu$.  We then vary
$\mu$ in the range of 100 GeV  $< \mu <$ 2 TeV that covers the typical
relic density satisfied higgsino mass zone of 1~TeV. In a correlated
scanning setup, for each value of $\mu$ we vary the common  
slepton mass parameter for all the three generations of sleptons  
within the range 50\% below and above the value of $\mu$. The common squark
mass parameters are chosen to be large (3 TeV).  
For squarks, we allow coannihilations only with the first two generations
for reasons described in Sec.\ref{sec2} while taking sleptons
as well as the third generation of squarks to be heavy (3~TeV).
As before, we vary
$\mu$ in the range of 100 GeV  $< \mu <$ 2 TeV. Then, 
for each value of $\mu$ we 
vary the common squark mass parameters for the first two generations 
within the range 50\% below and above the value of $\mu$. 

Furthermore, while $\mu$ being varied, we scan 
the trilinear soft breaking parameter $A_t$ from  
$-2$~TeV to $-7$~TeV, so as to satisfy the higgs mass data. However, we must
emphasize that the choice of $A_t$ has a very
small impact in our study of sfermion coannihilations.
The squark mass parameters of the third generation as well as the 
electroweak symmetry breaking  
(EWSB) scale are taken as 3~TeV throughout the slepton and squark
coannihilation studies.
The $SU(3)_C$ gaugino mass parameter $M_3$ 
is also chosen to be 3~TeV whereas the mass
of the CP-odd Higgs ($M_A$) is set at 5 TeV. The latter avoids
a Higgs resonance annihilation region. 
In order to study the effect of slepton coannihilations
on the relic density of DM,
we make sure that the physical slepton masses stay within 
20\% of the LSP mass irrespective of the generation.
The same is true for the
case of first two generations of squarks while we analyze the effects of squark
coannihilations.
We like to emphasize that with the above nearly 
degenerate squark masses close to that of the LSP, the commonly discussed 
LHC limits\cite{atlassquarklimitetc} for squarks would not apply to our scenario. 
\begin{figure}[!htb]
  \vspace*{0.3in}
  \vspace*{-0.05in}
  \mygraph{hfrac_slep}{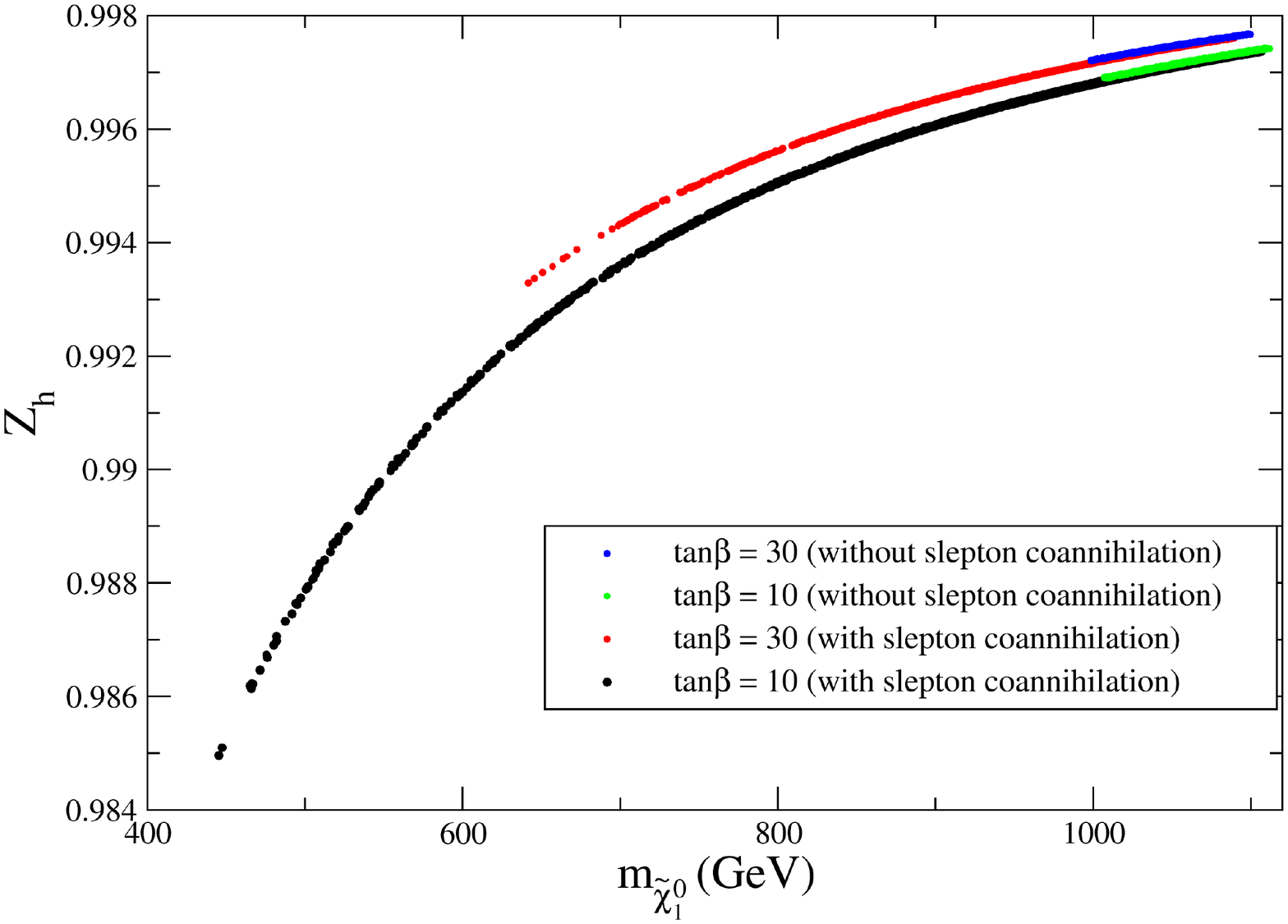}
  \hspace*{0.5in}
  \mygraph{hfrac_sq}{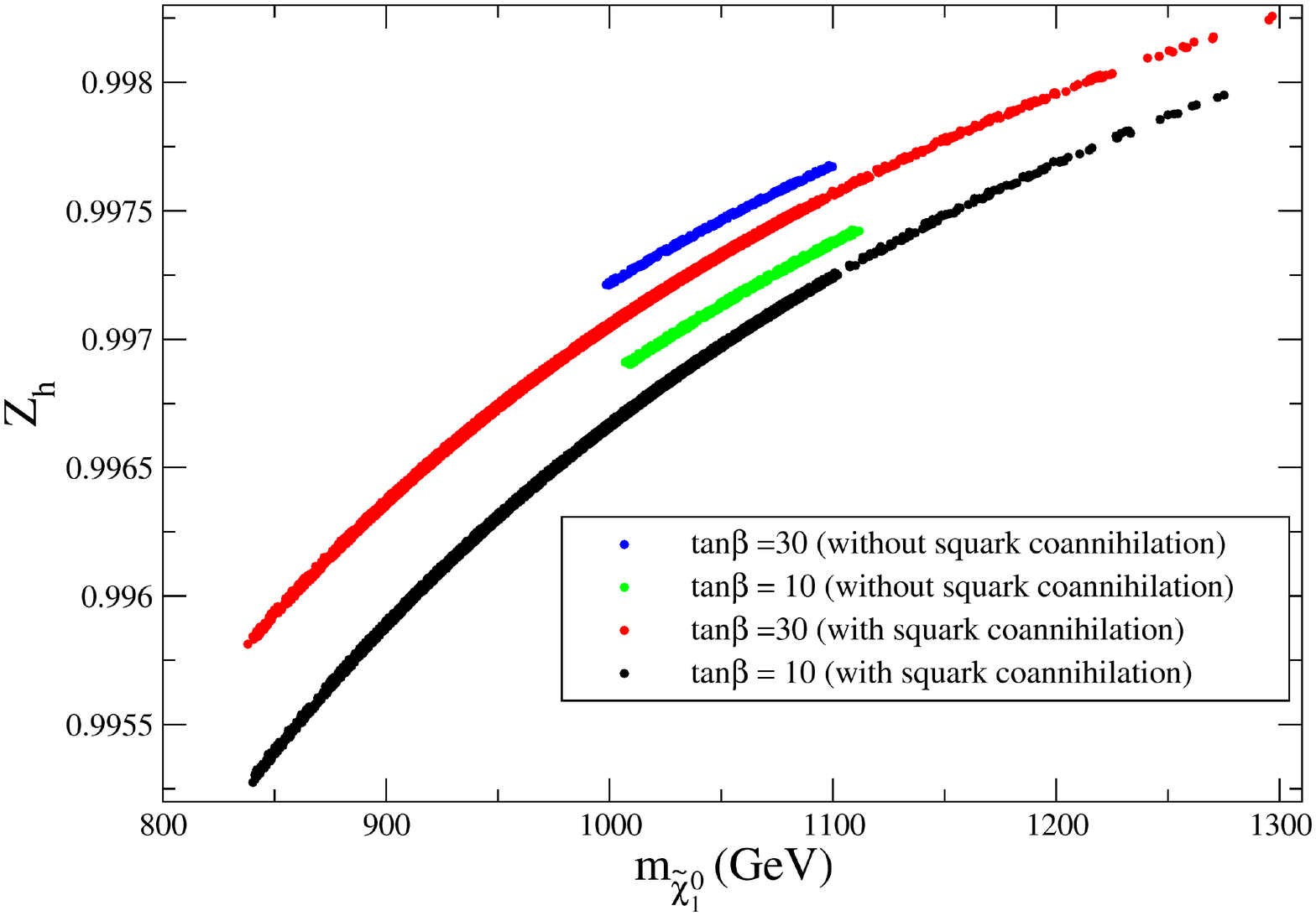}
  \caption{\small a) Plot of higgsino fraction $Z_h$ vs LSP mass for a higgsino
    dominated LSP when the LSP coannihilates with sleptons and sneutrinos
    of all the three generations apart from the usual LSP-$\chonepm$ and
    LSP-$\lsptwo$ coannihilations for $\tan\beta=10$ and $30$ as appropriate to the
    case of a higgsino-type of LSP. The plot is obtained by varying $\mu$ in a
    correlated higgsino-gaugino mass setup as explained in the text.
    The
    black and red points refer to $\tan\beta=10$ and $30$ respectively. 
    The reference results with no slepton coannihilations are shown in green and
    blue points for $\tan\beta=10$ and $30$ respectively. All the 
    points in the plot satisfy the lower and upper limits of DM
    relic density (Eq.\ref{planckdata}). Clearly, the above sfermion coannihilations reduce the
    lower limit of the LSP mass satisfying Eq.(\ref{planckdata}) to
    $\mlspone \simeq 450$~GeV from about 1~TeV. 
    b) Similar plot with LSP-squark coannihilations along with reference
    cases where squarks refer only to the first two
    generations. The color codes are same as those of (a).
    Reduction of the lower limit and enhancement of
    the upper limit of the LSP mass are notable.}
  \label{hfrac_caption}
\end{figure}

Fig.\ref{hfrac_slep} shows the scatter 
plot of the higgsino fraction $Z_h$ vs LSP mass when the LSP efficiently 
coannihilates with sleptons and sneutrinos of all the three generations.
We note that the LSP-sfermion coannihilations take place in the
background of strong electroweakino coannihilations.   
The significance of higgsino purity level in turn is directly
related to the level of coannihilations between the electroweakino states
$\lspone,\lsptwo$ and $\chonepm$.
The squarks are taken to be very heavy (3 TeV).
The reference results for the generic higgsino LSP or the case of no sfermion
coannihilations are shown in green and
    blue points for $\tan\beta=10$ and $30$ respectively.
  The 
  black and red points refer to $\tan\beta=10$ and $30$ respectively
  for the cases with slepton coannihilations. 
     All the 
    points in the scatter plot satisfy the lower and upper limits of DM
    relic density. Clearly as seen in Fig.\ref{hfrac_slep},
    the slepton coannihilations reduce the
    lower limit of the LSP mass satisfying Eq.\ref{planckdata} to  
    $\mlspone \simeq 450$~GeV for $\tan\beta=10$ and $\sim$~640~GeV
    for $\tan\beta=30$ respectively from about 1~TeV corresponding to the
    generic higgsino DM result. There is hardly any change in
    the upper
    limit of the mass of LSP in this regard. The difference of the lower limits for the two values of $\tan\beta$ for the case of
      slepton coannihilations arises from the L-R mixing of the
      third generation of sleptons (Sec.\ref{sec2}).
      A relatively larger spread of stau masses via exponential suppression
      effectively reduces the
      corresponding coannihilation contributions toward the
      effective cross section. 
      The cases of no slepton coannihilations
      hardly depend on $\tan\beta$. This is consistent with
      the discussion made in Sec.\ref{sec2}.
    We extend the results to LSP-squark coannihilations in Fig.\ref{hfrac_sq}.
    All the internal degrees of freedom are taken into account
    including the colors.
 The sleptons, on the other hand, are chosen to be very heavy (3 TeV)
 as mentioned before.
    The color codes are same as those of Fig.\ref{hfrac_slep}. 
    Here, squarks refer only to the first two
    generations for reasons mentioned in Sec.\ref{sec2}. For the lower limit 
    of the LSP mass, one finds $\mlspone \simeq 840$~GeV 
    irrespective of $\tan\beta$.    
 In any case, the above is rather a modest 
reduction from about 1~TeV corresponding to the 
    case of a generic higgsino DM. We must also note that for squark
    coannihilations the upper 
    limit of $\mlspone$ is stretched by about 200 GeV corresponding to the
    no sfermion coannihilation case.
In this zone of 
  large LSP mass, the relic density decreases compared to the generic
  higgsino DM case due to the
 dominance of squark-squark coannihilations. We will come back to it for
 further discussion while describing Fig.\ref{higgsino_delta_sq}. 

 \begin{figure}[!htb]
  \vspace*{0.3in}
  \vspace*{-0.05in}
  \mygraph{higgsino_delta_slep}{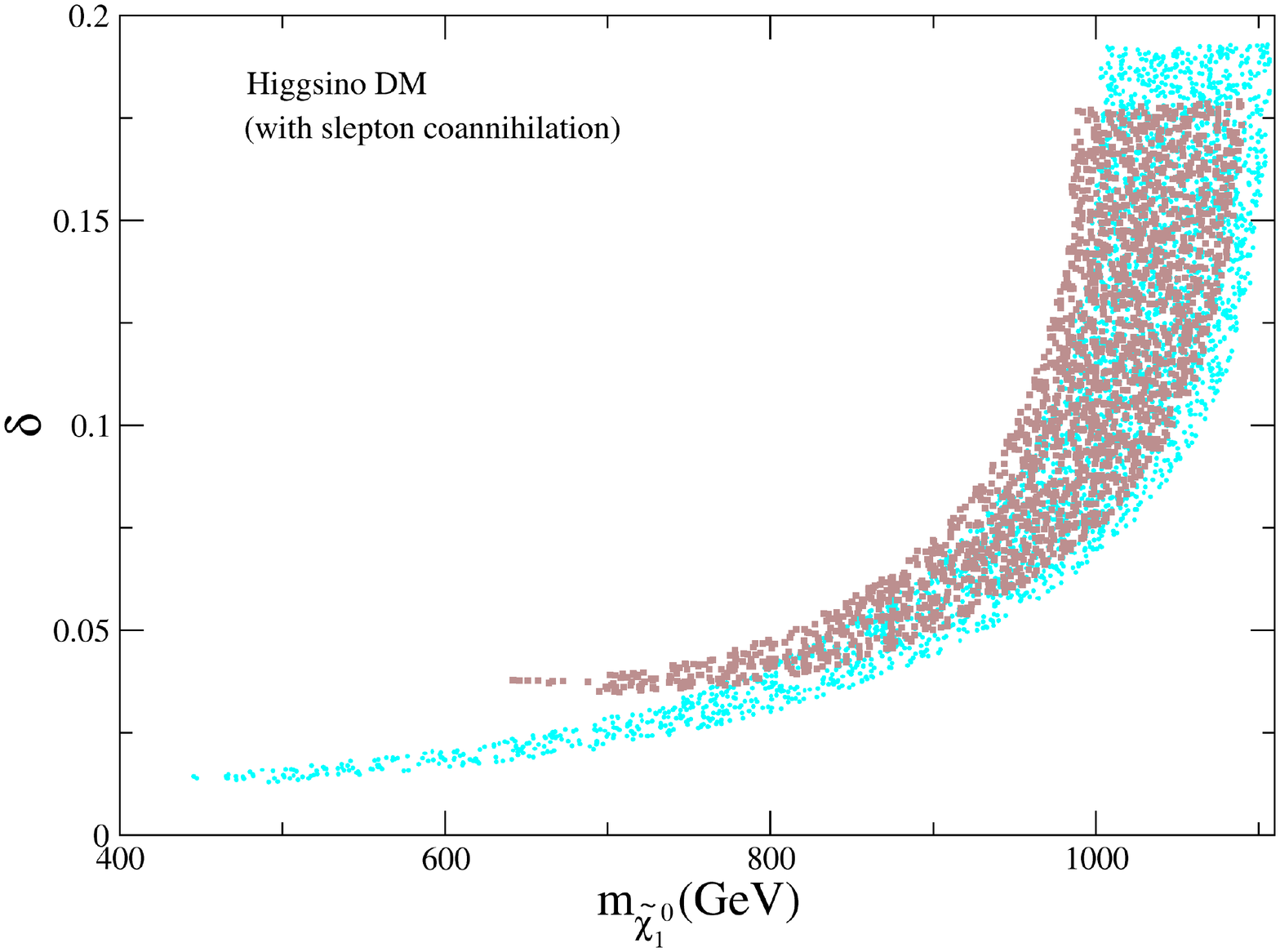}
  \hspace*{0.5in}
  \mygraph{higgsino_delta_sq}{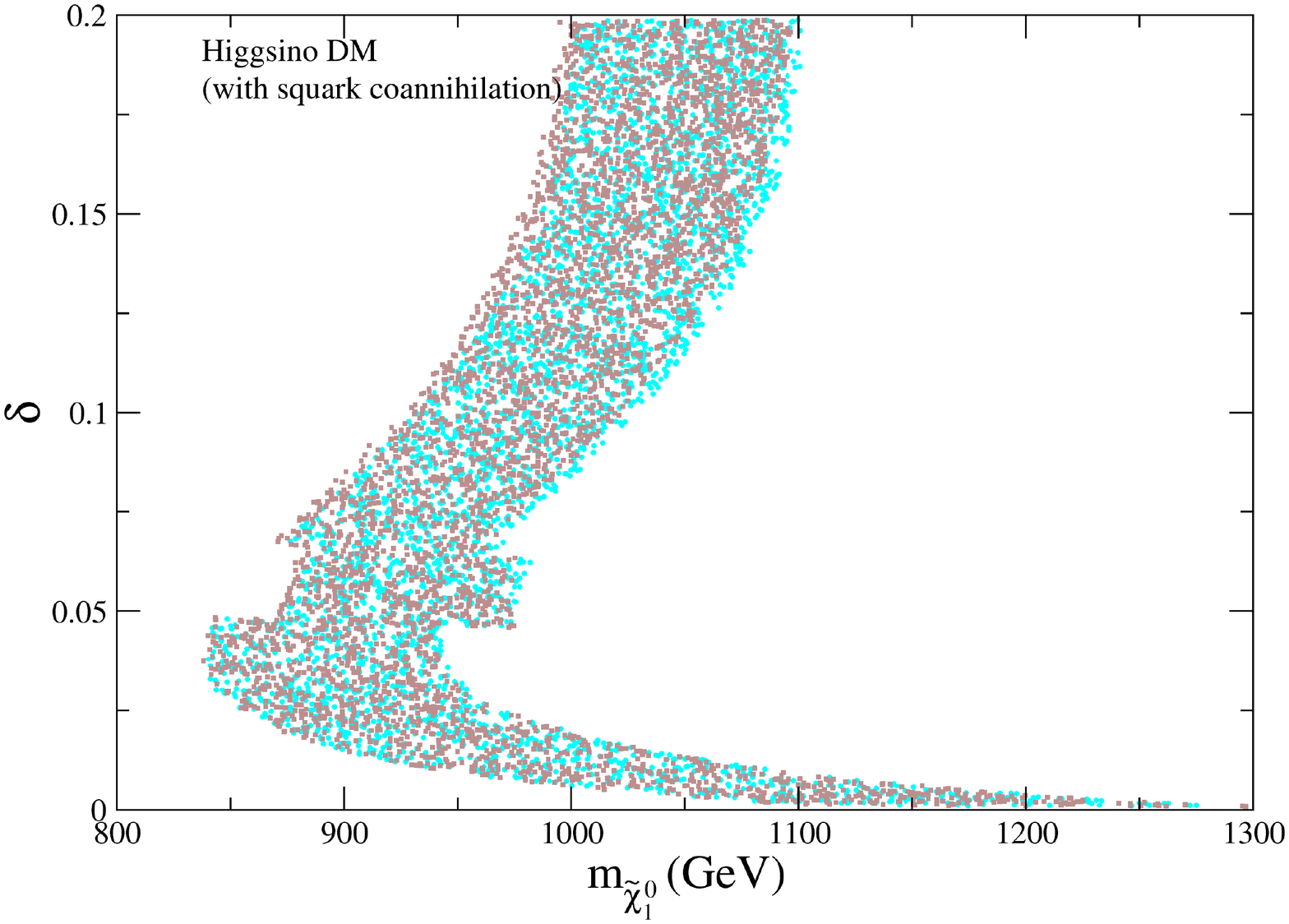}
\caption{\small a) Scatter plot for the mass of LSP $\mlspone$
    vs $\delta$,
    the relative deviation of
    slepton masses with respect to the mass of the LSP for the 
    LSP-slepton coannihilation scenarios including also sneutrinos, where
    the LSP is highly higgsino dominated in its composition.
    For a given
    slepton mass labeled as $m_{{\tilde l}_i}$ the relative deviation is given as
$\delta_i=\frac{(m_{{\tilde l}_i}-\mlspone)}{\mlspone}$.
    $\delta $ refers to the
    average of the values of $\delta_i$
    for all the three 
    generations of sleptons  including all the   
    internal degrees of freedom of sleptons as mentioned 
    in Sec.\ref{intro}. The two different colors namely cyan (circle)
    and brown (square)  
    refer to the cases of $\tan\beta=10$ and $30$ respectively. 
    b) The results of considering the coannihilations of LSP with the first
    two generations of squarks for $\tan\beta=10$ and $30$. The color scheme
    is similar to (a).}
\label{higgsino_delta_slepsq_caption}
\end{figure}

    Fig.\ref{higgsino_delta_slepsq_caption} shows scattered points in the
    $\mlspone-\delta$ plane corresponding to the analysis of Fig.\ref{hfrac_caption}.
Fig.\ref{higgsino_delta_slep} shows the scatter plot in the 
($\mlspone-\delta$) plane for LSP coannihilating with sleptons.
    The points that correspond to satisfying the DM relic
      density limits of Eq.\ref{planckdata} have two different colors namely cyan (circle)
    and brown (square)  
    representing the cases of $\tan\beta=10$ and $30$ respectively.
    The left
    and right hand side white regions indicate LSP to be
    underabundant and overabundant respectively. The regions with large
    $\delta$ are associated with smaller degrees of coannihilation.
    The brown region 
    corresponding to $\tan\beta=30$ has a larger degree of
    stau L-R mixing. 
This follows from the discussion made in Sec.\ref{sec2}.
Thus, compared to $\tan\beta=30$, the effect of
slepton coannihilations is more prominent, thereby meaning the lower limit 
of the LSP mass to become smaller for $\tan\beta=10$. As seen 
in the figure, this
    leads to a higgsino dominated LSP with mass as low as 450~GeV 
    satisfying Eq.\ref{planckdata}.
    The associated coannihilating sleptons 
    correspond to $\delta$ less than 2\%. We also note that as
      explained in Sec.\ref{sec2} as well as in the description of
      Fig.\ref{hfrac_slep}, an analysis with only the first two generations of sleptons
      would hardly show any dependence on $\tan\beta$ concerning the lower limit of the
      LSP mass satisfying the relic density constraint of
      Eq.\ref{planckdata}.   
    Fig.\ref{higgsino_delta_sq} shows the result for the LSP-squark
    coannihilations where we have considered only the first two generations of
    squarks for which the outcome is essentially independent of $\tan\beta$.
          On the lower side, the LSP mass satisfying Eq.\ref{planckdata}
      is reduced to around 840~GeV from 1~TeV corresponding to the generic
      higgsino DM case. 
On the other end, the higgsino LSP mass may extend to 
about 1300 GeV, about a 20\% increase than the generic higgsino DM
upper limit. There is a ``notch'' region
corresponding to $\delta \sim 0.05$ spreading across the values of
the LSP mass. In this quasi degenerate LSP-squark setup, 
the above arises due to a relatively 
rapid change of the DM relic density coming out of the enhancement of
$q {\tilde q} {\tilde \chi_{1,2}^0}$ effective couplings.
We will discuss this at the end of Sec.\ref{directDMsub}. 
    
        In the zone of relatively large
          LSP mass and for nearly
      degenerate squarks and the LSP {\it i.e.} small values of $\delta$,
          a detail check of the outgoing
      products of annihilation and coannihilations confirms that the  
      squark-squark coannihilations dominate  over the generic
      higgsino DM effective annihilation cross-section. We note that the  
      latter, which is inversely proportional to the DM relic density, 
      decreases with increase in higgsino mass\footnote{
        \label{dmrelivslspmass}
        For the generic higgsino LSP case one has
        $\Omega_{\lspone} h^2=
        0.10 { (\frac{\mu}{1~{\rm TeV}})}^2$\cite{arkani-hamed},
        where $\mu$ is given in TeV. A similar relation for a wino LSP 
      with mass $m_{\tilde W}$ reads $\Omega_{\tilde W} h^2=
      0.13 { (\frac{m_{\tilde W}}{2.5~{\rm TeV}})}^2=0.021
      m_{\tilde W}^2$\cite{arkani-hamed}, denoting a factor of 5 stronger effective annihilation cross section compared to the higgsino case. As we will
      see the squark-squark coannihilation contributions are not 
      large enough to supersede the generic wino DM depletion cross
      section.
      Hence, the
      wino dominated LSP scenario with squark coannihilations
      will not encounter any stretching of the LSP
      mass region satisfying the relic density data on the higher side.
      }.
      Additionally, for larger values of $\delta$ and larger LSP mass,
      in spite of a smaller degree of generic electroweakino
      annihilation/coannihilations due to heavier LSP,
      the squark-squark 
      coannihilations are more and more exponentially suppressed. 
      Thus, even for slightly larger values of $\delta$ we get
      overabundance of DM.

\begin{figure}[!htb]
\vspace*{0.3in}
\begin{center}
\includegraphics[scale=0.4]{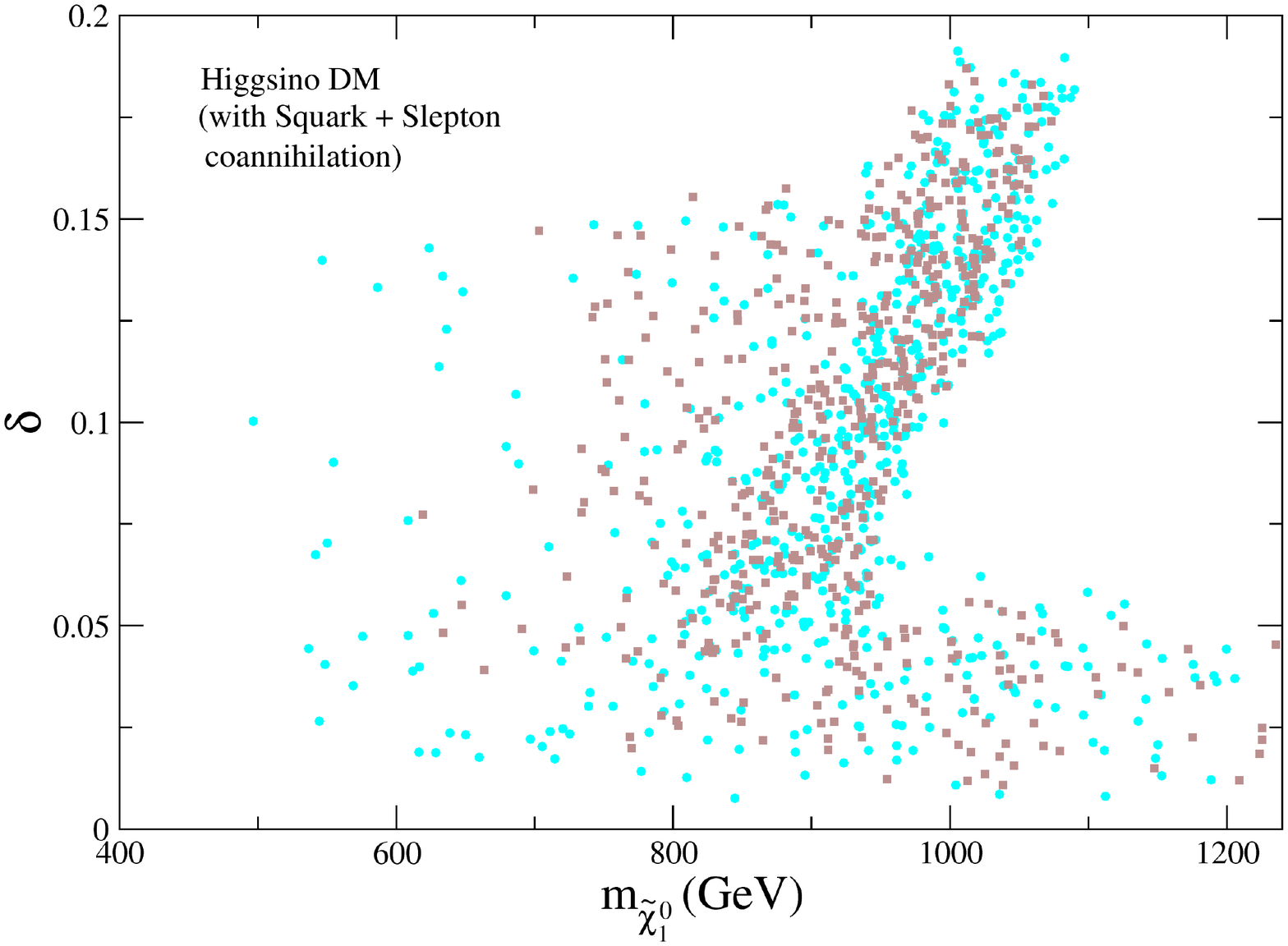}
\caption{\small The results of a combined analysis of
  slepton and squark coannihilations at 20\% level as mentioned in the text.
The cyan (circle) and brown (square) points
    refer to $\tan\beta=10$ and $30$ respectively.   All the 
    points in the scatter plot satisfy the lower and upper limits of DM
    relic density of Eq.(\ref{planckdata}).
  The lower limit of the higgsino LSP mass
satisfying the DM relic density constraint is around 500 GeV. 
}
\label{hsfermiontotal_mlsp}
\end{center}
\end{figure}      
The result of a combined analysis of
the above slepton (three generations) and squark coannihilations
(two generations) is given in
Fig.\ref{hsfermiontotal_mlsp}. The colors refer to the same convention as
that of Fig.\ref{higgsino_delta_slepsq_caption}. The top-squark masses are chosen to be very heavy (3 TeV).
The left 
    and right hand side blanck regions indicate LSP to be 
    underabundant and overabundant respectively. Here, a
      large number of sfermions participate in coannihilations and
      the region with small LSP mass that would have otherwise
      underabundant DM gets the right amount of relic density  
      even for relatively larger values of  
    $\delta$. The relic density is clearly enhanced thus  
      satisfying Eq.\ref{planckdata}. However, among a variety of participating
      coannihilation channels, 
      predominant contributions
    still come from electroweakino coannihilations.
 The lower limit of the higgsino LSP mass
satisfying the DM relic density constraint is around 500 GeV.

\subsection{Wino dominated LSP}
\label{subsec_wino}
\begin{figure}[!htb]
  \vspace*{0.3in}
  \vspace*{-0.05in}
  \mygraph{wfrac_slep}{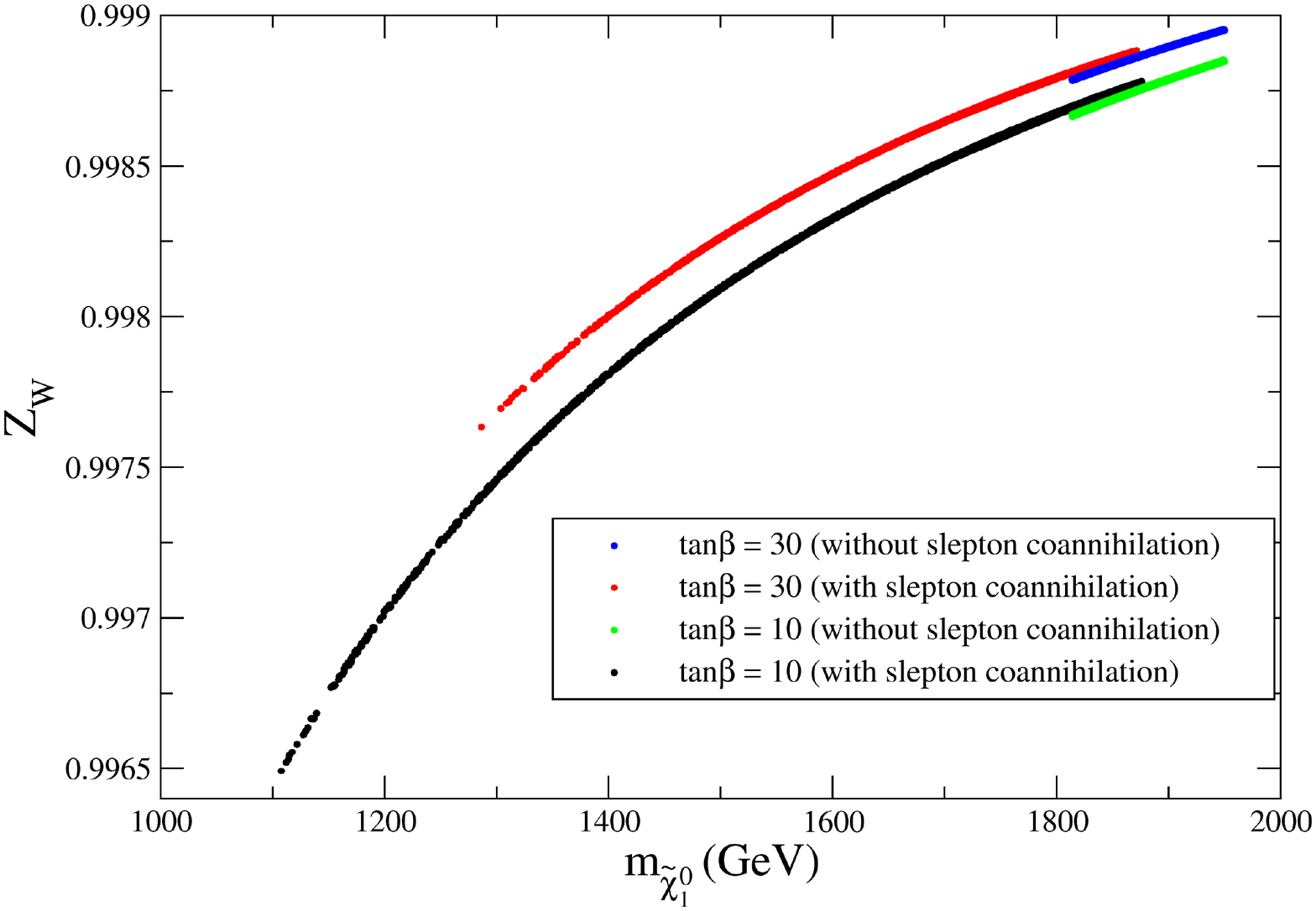}
  \hspace*{0.5in}
  \mygraph{wfrac_sq}{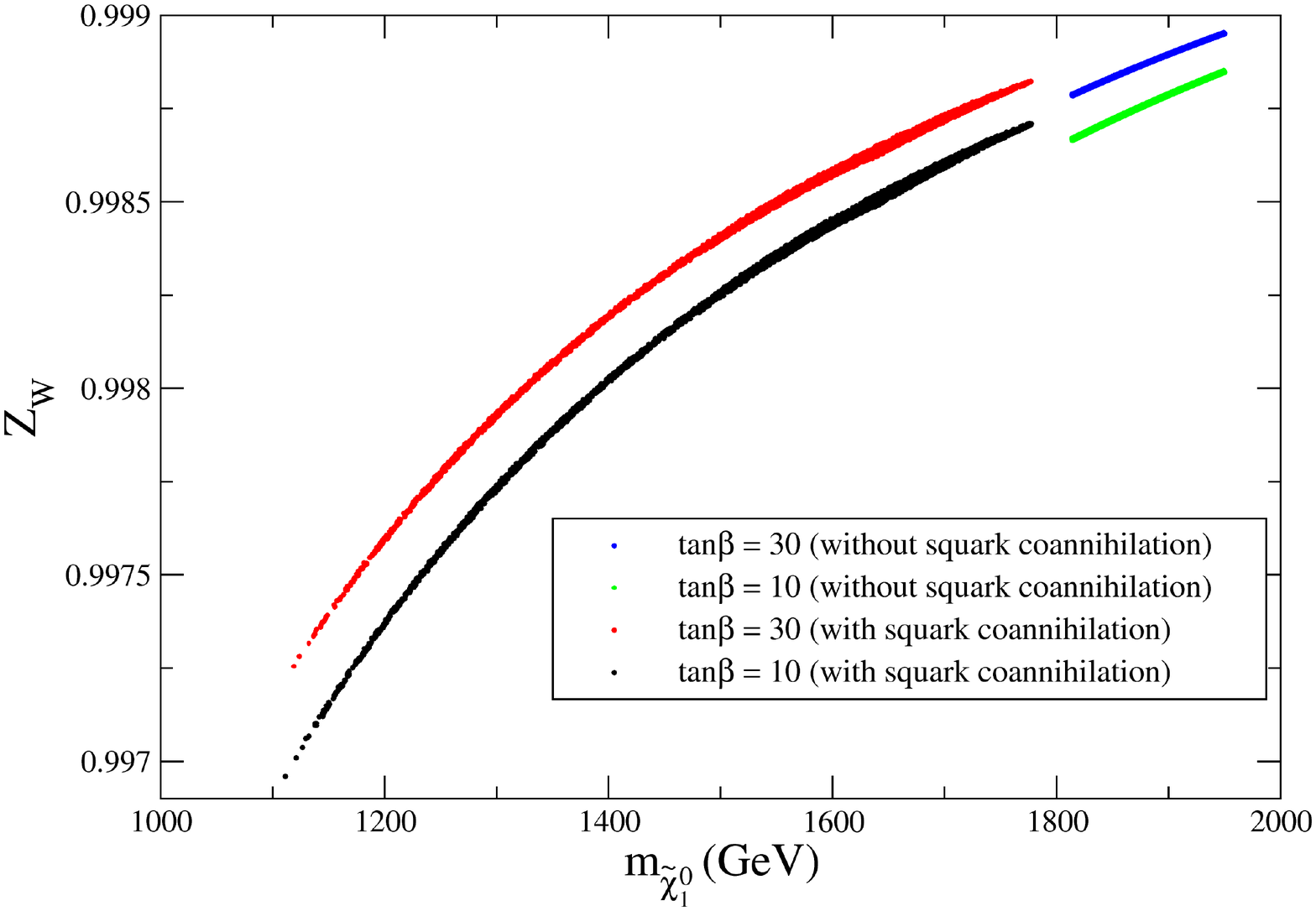}
  \caption{\small a) Plot of wino fraction $Z_W$
    vs LSP mass for a wino
    dominated LSP when the LSP coannihilates with sleptons and sneutrinos
    of all the three generations apart from the usual LSP-$\chonepm$
    coannihilations. The cyan (circle) and brown (square) points
    refer to $\tan\beta=10$ and $30$ respectively. All the 
    points that are generated by varying $M_2$
    satisfy the lower and upper limits of DM
    relic density of Eq.(\ref{planckdata}).
    Clearly, the slepton coannihilations reduce the
    lower limit of the LSP mass satisfying the WMAP/PLANCK data to
    $\mlspone \simeq 1.1$~TeV. 
    b) Similar plot with LSP-squark coannihilations along with reference cases with no LSP-squark coannihilations where squarks refer only to the first two
    generations. The color codes are same as those of (a).}
  \label{wfrac_caption}
\end{figure}

A wino-like LSP in MSSM implies nearly degenerate 
$\lspone$ and $\chonepm$, both of whose masses being essentially
determined by the SU(2) gaugino mass parameter $M_2$. 
The smallness of mass difference between $\mlspone$ and $\mchonepm$ 
leads to intense coannihilations resulting into a relic density
too low to satisfy the observed limits unless the mass of wino is too large
(above 2~TeV). A choice like $M_1> \mu >M_2$ would give rise to a
wino dominated LSP. In particular, we choose  
$\mu = 2 M_2$ and $M_1 > 2.4 M_2$, so as to make the LSP
predominantly a wino. $M_2$ is then
varied in the range 100 GeV $< M_2 <$ 2.5 TeV. 
For slepton coannihilations and a given value of $M_2$,
the common mass parameter for 
the slepton masses of all the three generations are varied within the range 
$50 \%$ below and above the value of $M_2$. The common squark mass parameters
are pushed to 4 TeV. 
On the other hand, for squark coannihilations, the common squark mass
parameter of the first two 
generations are similarly chosen around the value of $M_2$, while the latter 
being scanned as before. Here, the slepton mass parameters are large (4 TeV). 
 The squark masses of the third
generation and the SU(3) gaugino mass parameter $M_3$ are kept at 4 TeV
while the CP-odd Higgs mass is set at 6 TeV throughout our analysis, 
thus ensuring no s-channel Higgs resonance annihilations. 
Once again, owing to the variation of $M_2$ that results into varying $\mu$ 
we scan $A_t$ between $-2$~TeV to $-7$~TeV so as to have the higgs mass
$m_h$ in the correct range.
It can be observed from
Fig.\ref{wfrac_caption} that in the absence of any slepton coannihilation 
the relic density becomes viable for 
$\mlspone \sim$ 1.8 TeV
\footnote{We must note that a wino mass of
1.8 TeV
  satisfying the relic density data is low compared to what
  is seen in the literature, typically 
  above 2~TeV.
  Obtaining a heavier wino that satisfies the DM relic density limits
  is possible 
  i) via considering larger sfermion mass and $\mu$ as explained in the text 
  in Sec.\ref{massdiff} and ii) most importantly, 
  via including non-perturbative 
  effects like Sommerfeld correction. Sommerfeld correction is 
  known
  to increase the wino mass that satisfies the relic density limits. We have
  not included such an effect particularly for the fact that a low wino mass like
  1.1~TeV would hardly have an appreciable degree of Sommerfeld effect. We
  would like to mention Ref.\cite{Beneke:2016ync} (their Fig.2)
  and Ref.\cite{Cirelli:2007xd} (their Fig.2) in support of the
  smallness of the correction for our relevant zone of wino mass. Considering
  the fact that the relic density $\propto M_2^2$, using
  Ref.\cite{Cirelli:2007xd} we estimate a 10-12 \% level of enhancement of
  $M_2$ for its lower bound that would satisfy the observational
  relic density limits. Thus, the lower limit of the wino mass is estimated 
  to change from 1.1 TeV to around 1.2 TeV (as we will come across
  in Fig.\ref{wfrac_caption}) if we include the Sommerfeld
  effect. 
  
}.
The presence of sleptons with masses close to $\mlspone$ 
leads to many new coannihilation channels and affects the
averaging procedure toward the effective cross section and as we will see this 
increases the DM relic density so that Eq.\ref{planckdata} is satisfied 
for much smaller masses of the LSP. 

\begin{figure}[!htb]
  \vspace*{0.3in}
  \vspace*{-0.05in}
  \mygraph{wino_delta_slep}{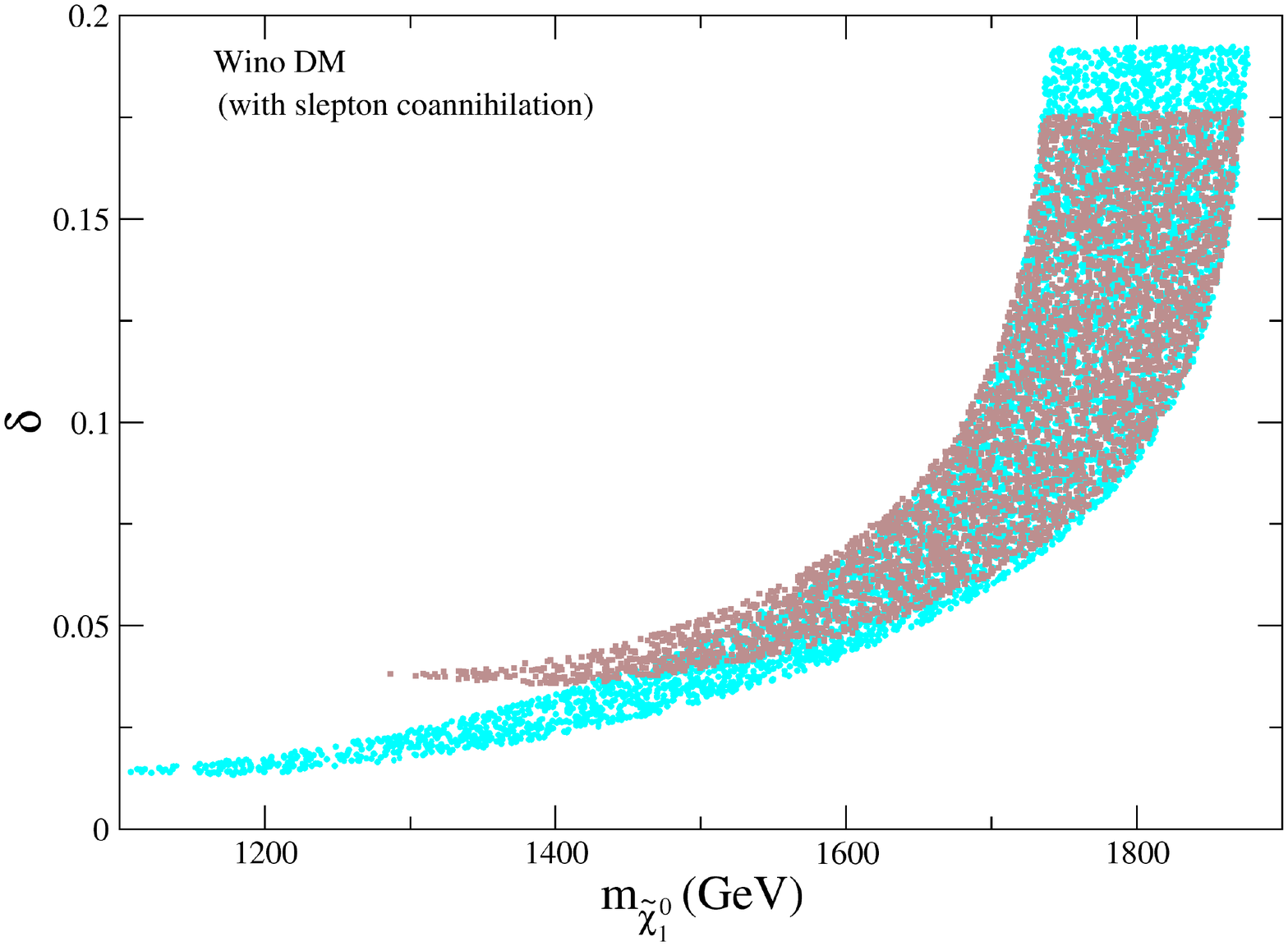}
  \hspace*{0.5in}
  \mygraph{wino_delta_sq}{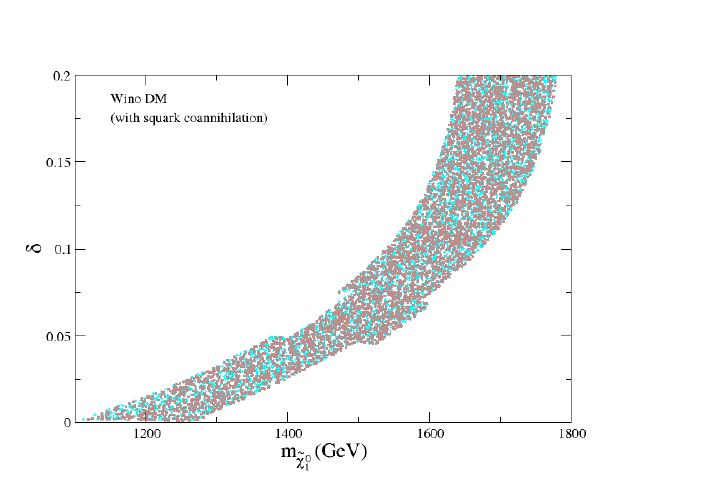}
\caption{\small a) Scatter plot for the mass of LSP $\mlspone$
    vs $\delta$,
    the relative deviation of
    slepton masses with respect to the mass of the LSP for
    LSP-slepton coannihilation scenarios including also sneutrinos, where
    the LSP is highly wino dominated in its composition.
    For a given
    slepton mass labeled as $m_{{\tilde l}_i}$ the relative deviation is given as
$\delta_i=\frac{(m_{{\tilde l}_i}-\mlspone)}{\mlspone}$.
    $\delta $ refers to the
    average of the values of $\delta_i$
    for all the three 
    generations of sleptons  including all the   
    internal degrees of freedom of sleptons as mentioned 
    in Sec.\ref{intro}. The two different colors namely cyan (circle)
    and brown (square)  
    refer to the cases of $\tan\beta=10$ and $30$ respectively. 
    b) The results of considering the coannihilations of LSP with the first
    two generations of squarks for $\tan\beta=10$ and $30$. The color scheme
    is similar to (a).}
  \label{wino_delta_slepsq_caption}
\end{figure}

\begin{figure}[!htb]
\vspace*{0.3in}
\begin{center}
\includegraphics[scale=0.4]{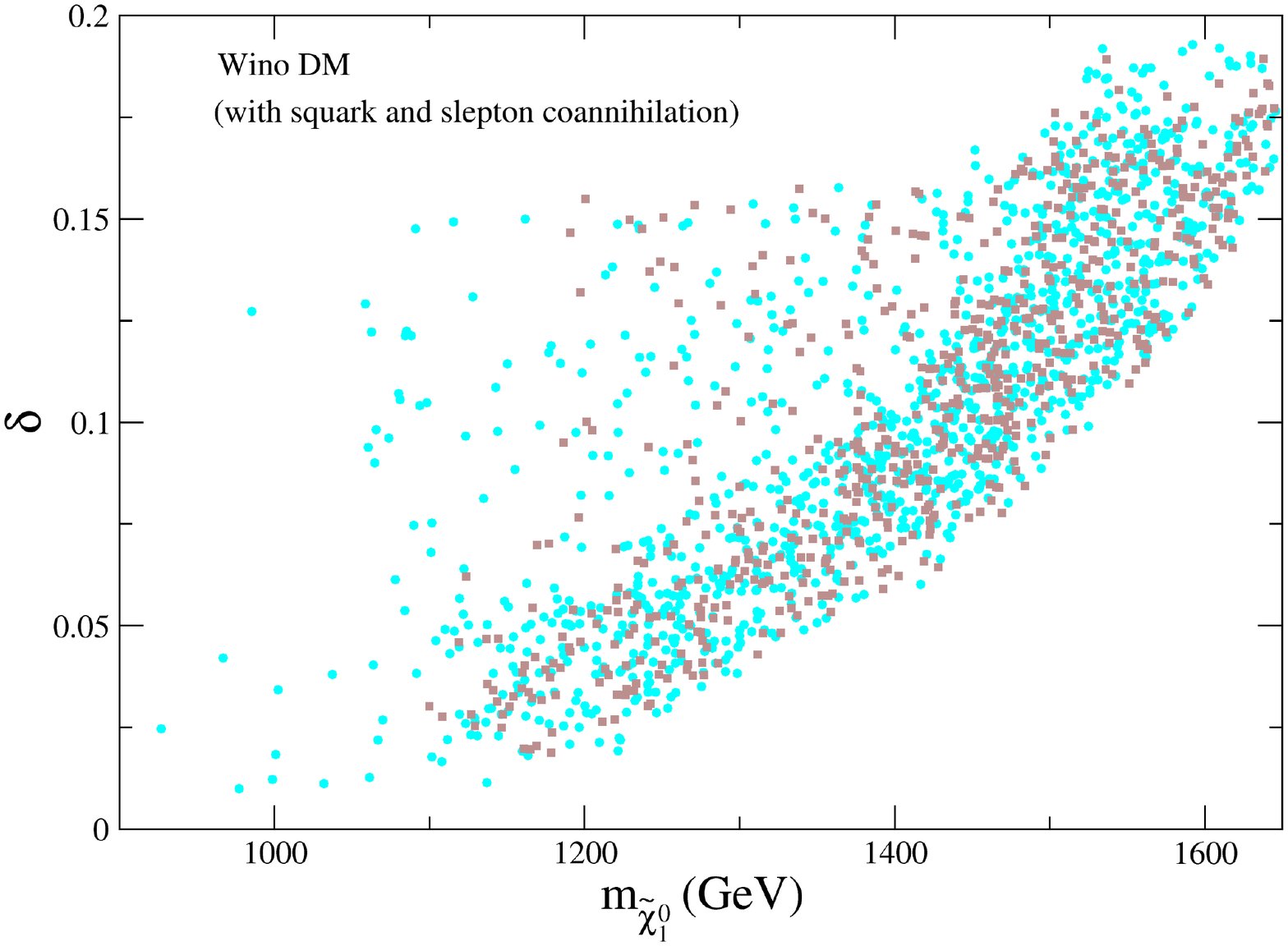}
\caption{\small The results of a combined analysis of
  slepton and squark coannihilations at 20\% level as mentioned in the text.
The cyan (circle) and brown (square) points
    refer to $\tan\beta=10$ and $30$ respectively.   All the 
    points in the scatter plot satisfy the lower and upper limits of DM
    relic density of Eq.(\ref{planckdata}).
  The lower limit of the wino LSP mass
satisfying the DM relic density constraint is around 900~GeV. 
}
\label{wsfermiontotal_mlsp}
\end{center}
\end{figure}

The amount of the wino component in the LSP is expressed in terms
of the wino fraction defined as, $Z_W = N_{12}^{\rm 2}$.
The wino fractions for 
different LSP masses with/without sfermion coannihilations are shown in
Fig.\ref{wfrac_caption}. We only show 
the parameter points that satisfy Eq.\ref{planckdata} for the limits of DM relic density.
Fig.\ref{wfrac_slep} shows the scatter
plot of the wino fraction $Z_W$ vs LSP mass when the LSP efficiently 
coannihilates with sleptons and sneutrinos of all the three generations.
We note that the LSP-sfermion coannihilations take place in the
background of strong electroweakino coannihilations.   
The significance of wino purity level in turn is directly
related to the level of coannihilations between the electroweakino states 
$\lspone$ and $\chonepm$.
As before, we have considered a maximum of 20\% deviation in masses 
for the coannihilating particles with respect to the mass of the LSP.
The squark masses of the first two generations 
are kept at 4 TeV .
The reference results with no slepton coannihilations are shown in green and
    blue points for $\tan\beta=10$ and $30$ respectively.
  The 
  black and red points refer to $\tan\beta=10$ and $30$ respectively
  for the cases with slepton coannihilations. 
   Clearly, as seen in Fig.\ref{wfrac_slep},
    the slepton coannihilations reduce the 
    lower limit of the LSP mass satisfying Eq.\ref{planckdata} to 
    $\mlspone \simeq 1.1$~TeV and 1.3~TeV respectively for 
$\tan\beta=10$ and $30$.
There is a dissimilarity in the lower limits in the results for the two 
different values of $\tan\beta$ for similar reasons as in the case of
higggsino LSP mentioned in Sec.\ref{subsec_higgsino}.
We extend the results to LSP-squark coannihilations in Fig.\ref{wfrac_sq}.
Here the sleptons are chosen to be very heavy (4 TeV).
    The color codes are same as those of Fig.\ref{wfrac_slep}. 
    The squarks again refer only to the first two
    generations for reasons mentioned earlier. 
    One finds the lower limit as $\mlspone \simeq 1.1$~TeV for 
both values of $\tan\beta$.

Fig.\ref{wino_delta_slepsq_caption} shows the average degree of
mass difference among the coannihilating partners while 
considering the coannihilations of LSP separately with 
sleptons or squarks where the LSP is wino dominated in its composition.
This refers to the scanning 
corresponding to Fig.\ref{wfrac_caption}.
The colored points belong to parameter space that satisfy
Eq.\ref{planckdata}. 
Fig.\ref{wino_delta_slep} shows the scatter plot
in the ($\mlspone$-$\delta$) plane. 
Apart from the  
    sleptons we also 
    include the three generations of sneutrinos in this analysis.
    $\delta$ is similarly defined as in the higgsino case of Sec.\ref{subsec_higgsino}.    
    The color codes are same as those of Fig.\ref{higgsino_delta_slepsq_caption}. 
    The left and right hand side white regions indicate LSP to be
    underabundant and overabundant respectively. The regions with large
    $\delta$ refer to smaller degrees of coannihilation because of
    larger exponential suppression. The brown region
    corresponding to $\tan\beta=30$ is associated with a larger degree of
    stau L-R mixing. 
Demanding both ${\tilde \tau}_1$ and
    ${\tilde \tau}_2$ along with the first 
    two generations of sleptons to have masses within 20\% of the LSP mass
    restricts the 
reach of $\delta$ for reasons mentioned in Sec.\ref{sec2}.
The effect of slepton coannihilations is 
more prominent for $\tan\beta=10$ and this
    leads to a wino dominated LSP having the right abundance with mass as low as 1.1~TeV
    when the coannihilating sleptons have $\delta$ less than 2\%. Similar to
    the higgsino analysis, the
    $\tan\beta$ dependence of the lower limit of $\mlspone$
    satisfying Eq.\ref{planckdata} would cease
    to exist if we had excluded the third generation of sleptons to take part
    in coannihilations.  
        Fig.\ref{wino_delta_sq} shows a similar result for the LSP-squark
    coannihilations where we consider only the first two generations of
    squarks as before. Clearly, being devoid of any top-squark coannihilations
    the result is essentially independent of
    $\tan\beta$. The lowest LSP mass that satisfies the DM relic density
    constraint is around 1.1~TeV. Unlike the higgsino case,
      there is no dominance of
    squark-squark coannihilations over the parameter space that satisfies the
    DM relic density constraint. This is indeed related to the large
    annihilation cross section that a wino has compared to that of a 
    higgsino for a given mass of the LSP
    (see footnote\#\ref{dmrelivslspmass}).
    Similar to the higgsino case, there is a ``notch'' region
corresponding to $\delta \sim 0.05$ spreading across the values of
the LSP mass. In this quasi degenerate LSP-squark setup, 
the above arises due to a relatively  
rapid change of the DM relic density coming out of the enhancement of
$q {\tilde q} {\tilde \chi_{1}^0}$ effective coupling.
We will discuss this at the end of Sec.\ref{directDMsub}.

    The results of a combined analysis of
the above slepton (three generations) and squark coannihilations
(two generations) is given in
Fig.\ref{wsfermiontotal_mlsp}.
    The two different colors namely cyan (circle) 
    and brown (square)  
    refer to the cases of $\tan\beta=10$ and $30$ respectively. The left
    and right hand side white regions indicate LSP to be
    underabundant and overabundant respectively. The regions with large
    $\delta$ refer to smaller degrees of coannihilation.
The lower limit of the wino LSP mass
satisfying the DM relic density constraint is around 900 GeV.

\noindent
We would like to mention here that the ATLAS and CMS collaborations
presented their results for chargino searches in the high transverse
momentum ($p_T$) disappearing tracks \cite{disap_track} and long lived
particle search channels for nearly
degenerate $\chonepm$ and $\lspone$\cite{llp}. 
The mass range of $\chonepm$ considered in this analysis is well within 
these bounds.

\section{Direct and Indirect Detection of DM}
\label{detectionDM}
In this section we will probe the prospect of direct and indirect detection
of the lightest neutralino. We will particularly
come across the importance of squark exchange diagrams in computing the
SI direct detection cross section.
The squark exchange diagrams are usually less important since the Higgs exchange
diagrams typically dominate. 
As we will see, in this analysis we are in a different situation
because of considering
quasi-degenerate squarks and LSP for the requirement of
coannihilations.

\subsection{Direct Detection}
\label{directDMsub}
Direct detection of DM involves finding the recoil energy
deposited when a DM particle 
scatters off a detector nucleus\cite{Kamionkowski,Silk}.
Spin-independent LSP-proton scattering may take place through s-channel squark exchange and t-channel Higgs exchange diagrams.  
When the squarks are considerably heavy, the Higgs exchange 
diagrams typically dominate\cite{Drees:1993bu,Chattopadhyay:2010vp}\footnote{On the contrary,  we will soon discuss the scenario when the
  squark exchange diagrams may even dominate over the Higgs exchange diagrams.}. 

\noindent
The Higgs-$\lspone$-$\lspone$ coupling for the higgsino-LSP case 
can be written down in the form\cite{hisano_nojiri}:
\begin{eqnarray}
C_{h \tilde \chi \tilde \chi} & \simeq & \mp \frac{1}{2} M_Z c_W
\bigl[ 1 \pm \sin2\beta \bigr]
    \biggl[ \frac{t_W^2}{M_1 - |\mu|}  + \frac{1}{M_2 - |\mu|} \biggr], \nonumber \\
C_{H \tilde \chi \tilde \chi} & \simeq & \frac{1}{2} M_Zc_W \cos2\beta
    \biggl[ \frac{t_W^2}{M_1 - |\mu|}  + \frac{1}{M_2 - |\mu|} \biggr] ,
\label{hSIcoup}
\end{eqnarray}
\noindent
where $t_W=\tan\theta_W$ etc. with $\theta_W$ being the Weinberg angle. 
Similarly, for the wino-LSP case, the couplings are as follows\cite{hisano_nojiri}:
\begin{eqnarray}
    C_{h \tilde \chi \tilde \chi} & \simeq & \frac{M_Z c_W}{M_2^2 - \mu^2}
    \bigl[ M_2 + \mu \sin2 \beta \bigr], \nonumber \\
    C_{H \tilde \chi \tilde \chi} & \simeq & - \frac{M_Z c_W}{M_2^2 - \mu^2}
    \mu \cos2\beta .
\label{wSIcoup}
\end{eqnarray}

\begin{figure}[!htb]
  \vspace*{0.3in}
  \vspace*{-0.05in}
  \mygraph{higgsino_si_slep}{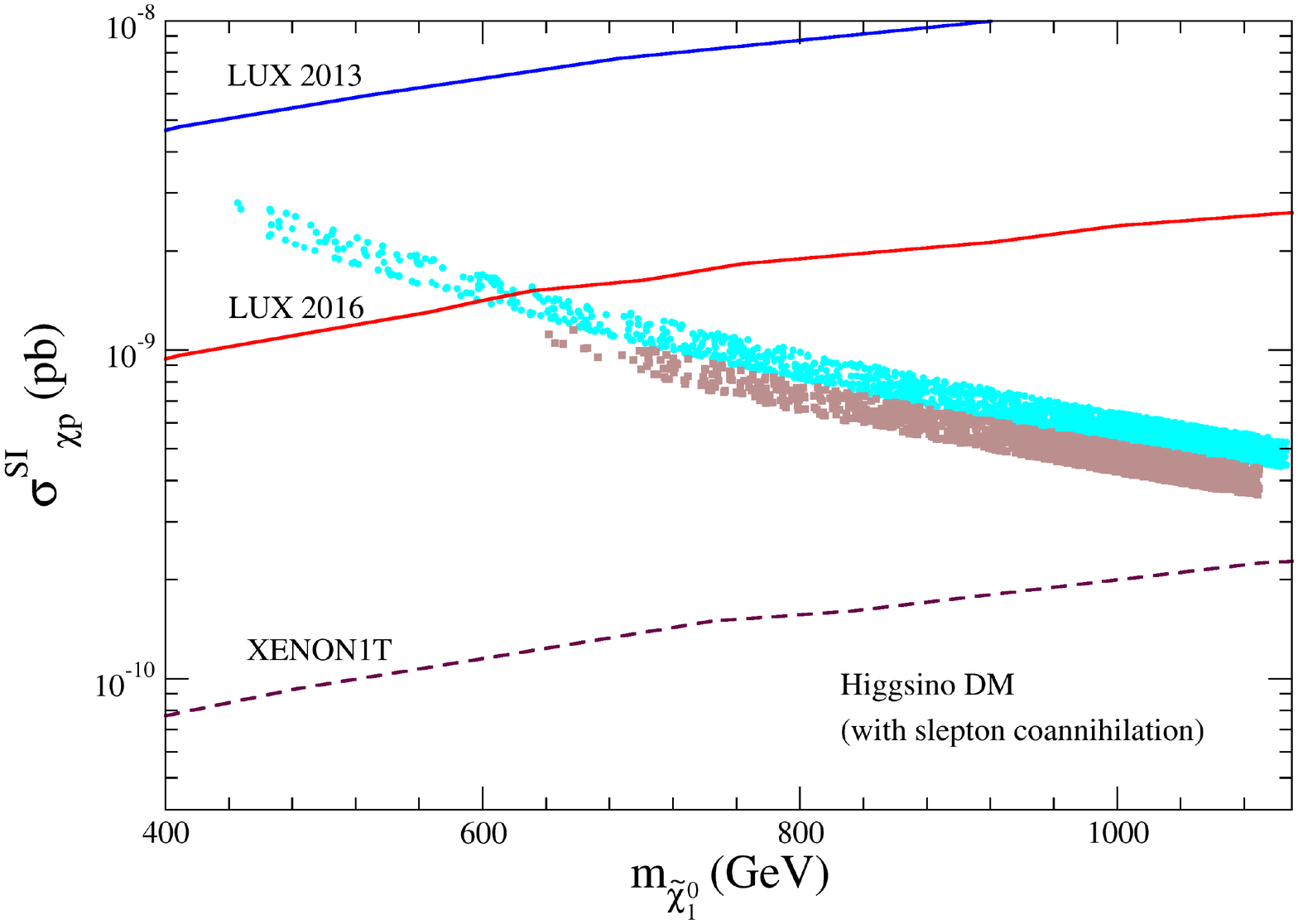}
  \hspace*{0.5in}
  \mygraph{higgsino_si_sq}{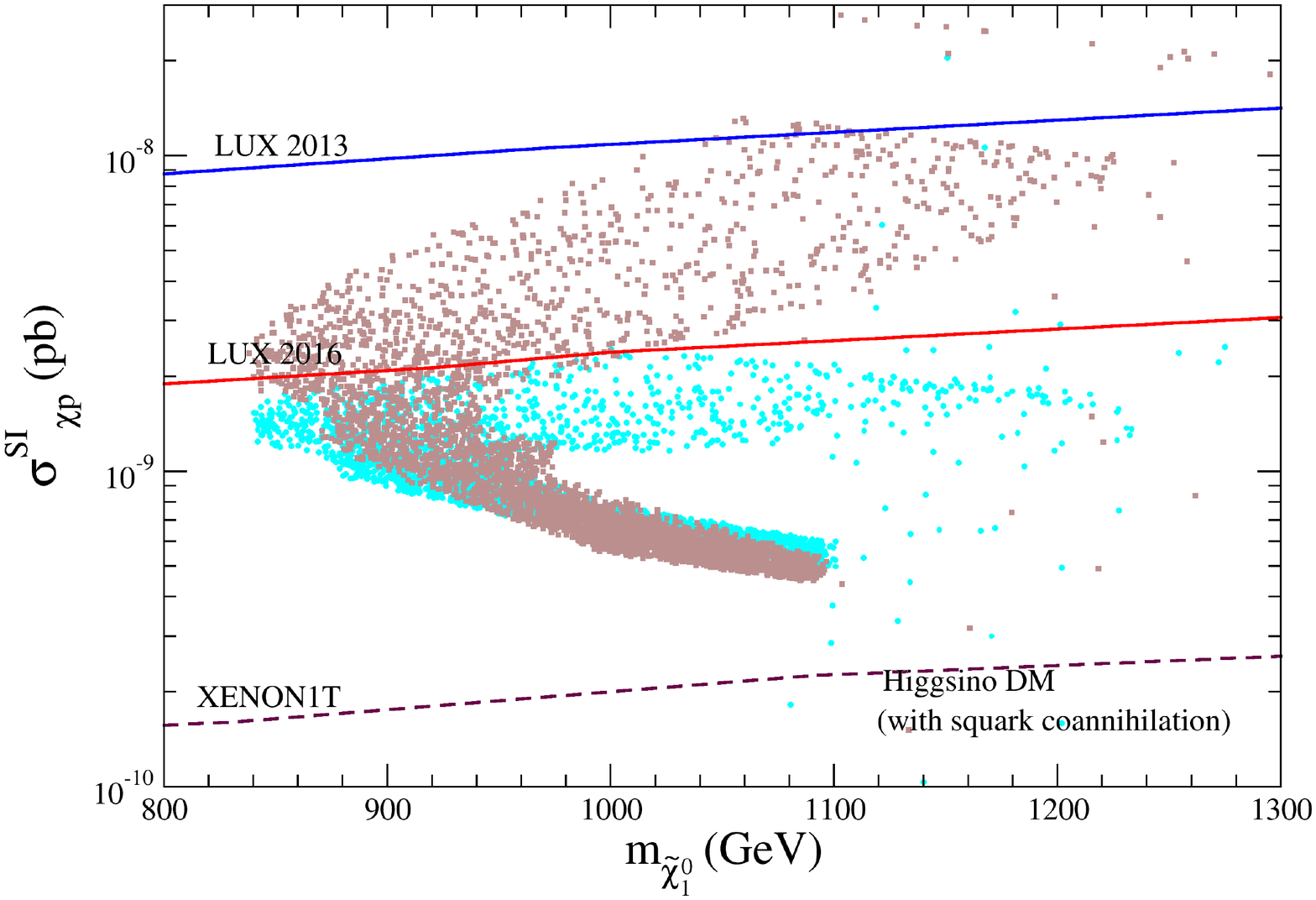}\\
   \begin{center}
  \mygraphthree{higgsino_si_slep_sq}{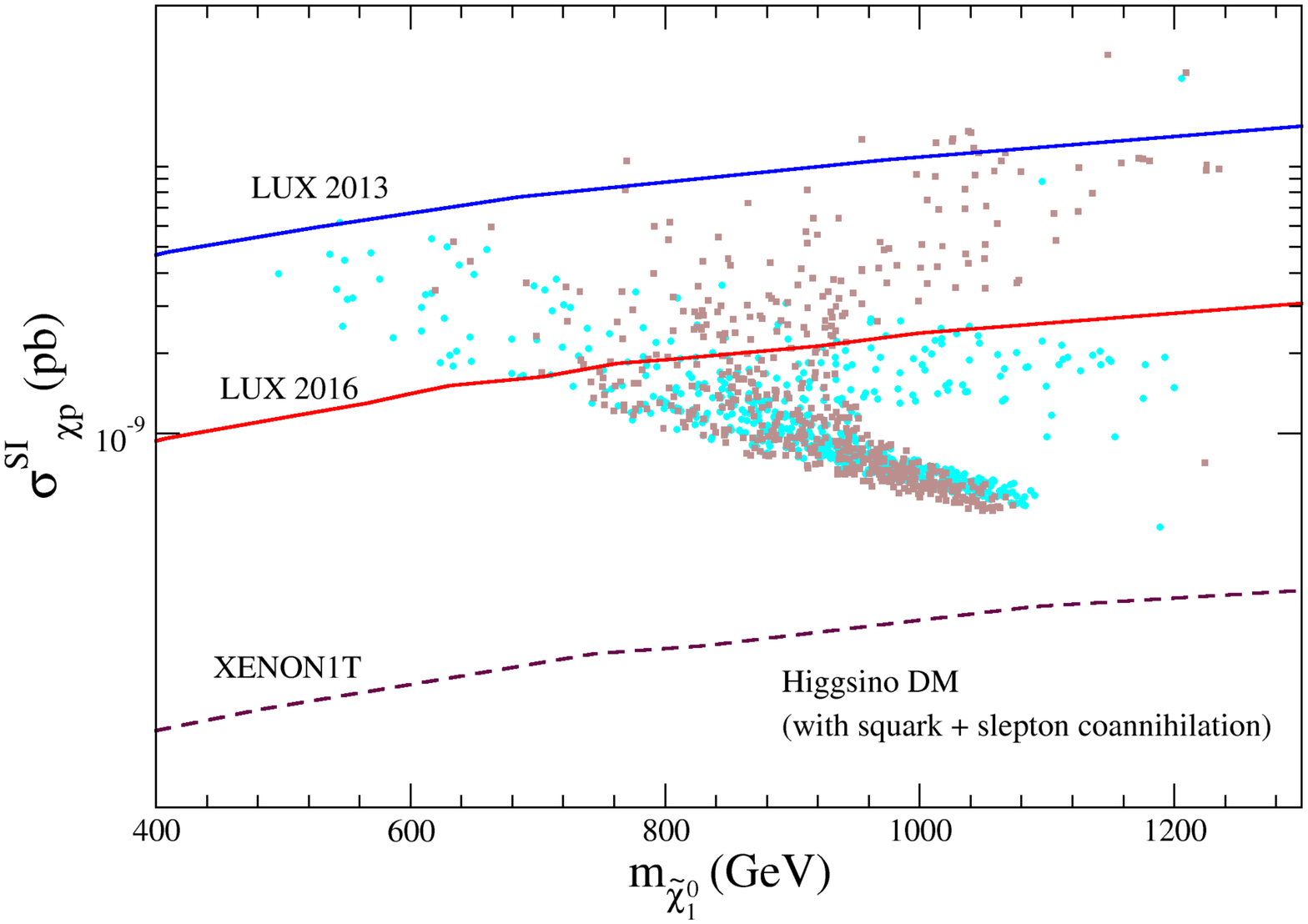}
  \end{center}
  \caption{\small (a) Scatter plot of spin-independent DM 
direct detection cross-section
    vs LSP mass 
      for higgsino dominated LSPs undergoing   
      LSP-slepton coannihilations. 
The cyan and brown points represent tan$\beta=$ 10 and 30 respectively 
 that satisfy Eq.(\ref{planckdata}). The red line (solid) is the LUX 2016
      exclusion 
contour and the maroon dashed line shows the expected limit from the future
XENON1T experiment. (b) Same as (a) except LSP undergoing LSP-squark 
coannihilations. (c) Same as (a) except LSP undergoing slepton plus squark 
coannihilations.
}
  \label{hsi_slep_sq_caption}
\end{figure}

\begin{figure}[!htb]
  \vspace*{0.3in}
  \vspace*{-0.05in}
  \mygraph{wino_si_slep}{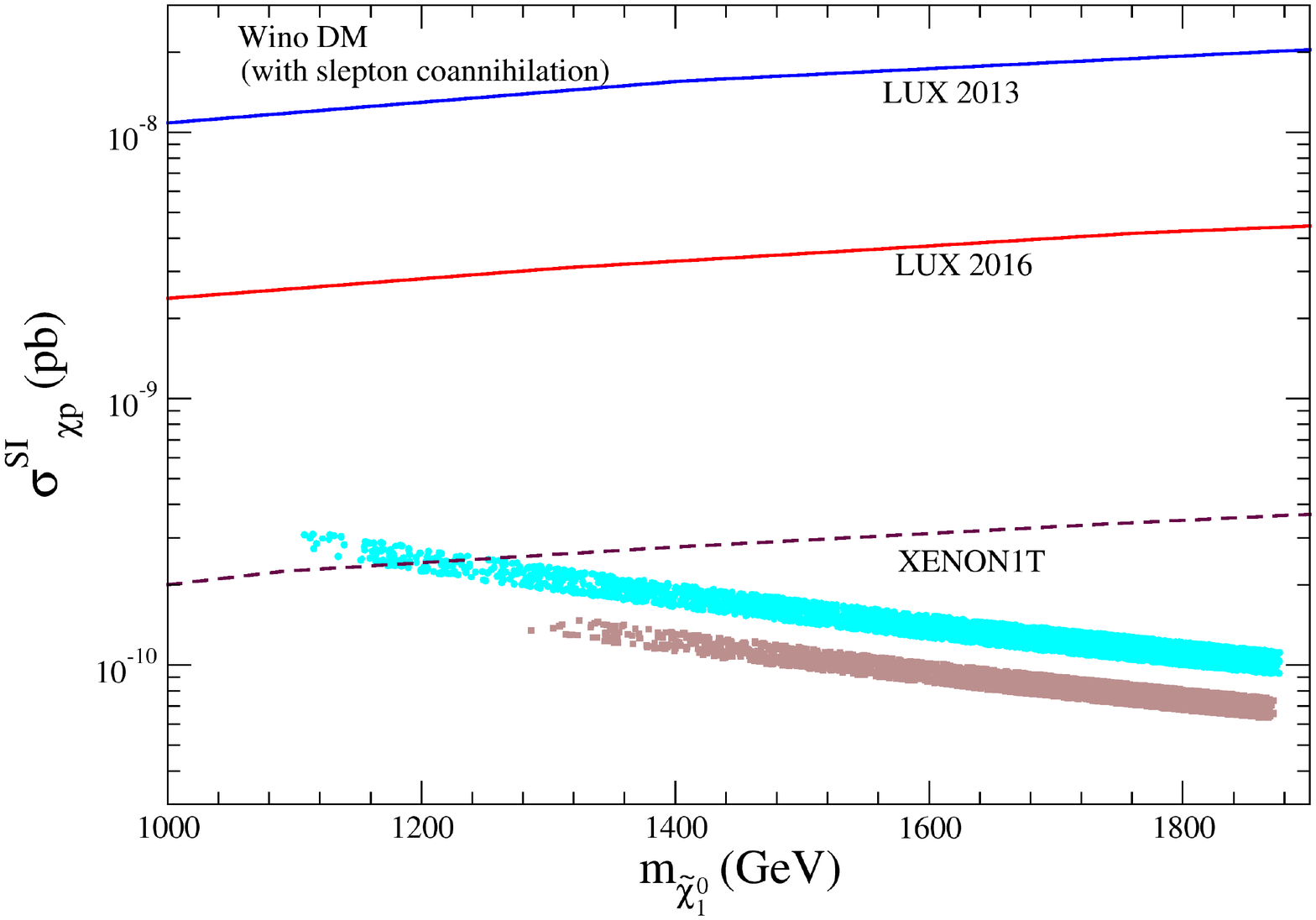}
  \hspace*{0.5in}
  \mygraph{wino_si_sq}{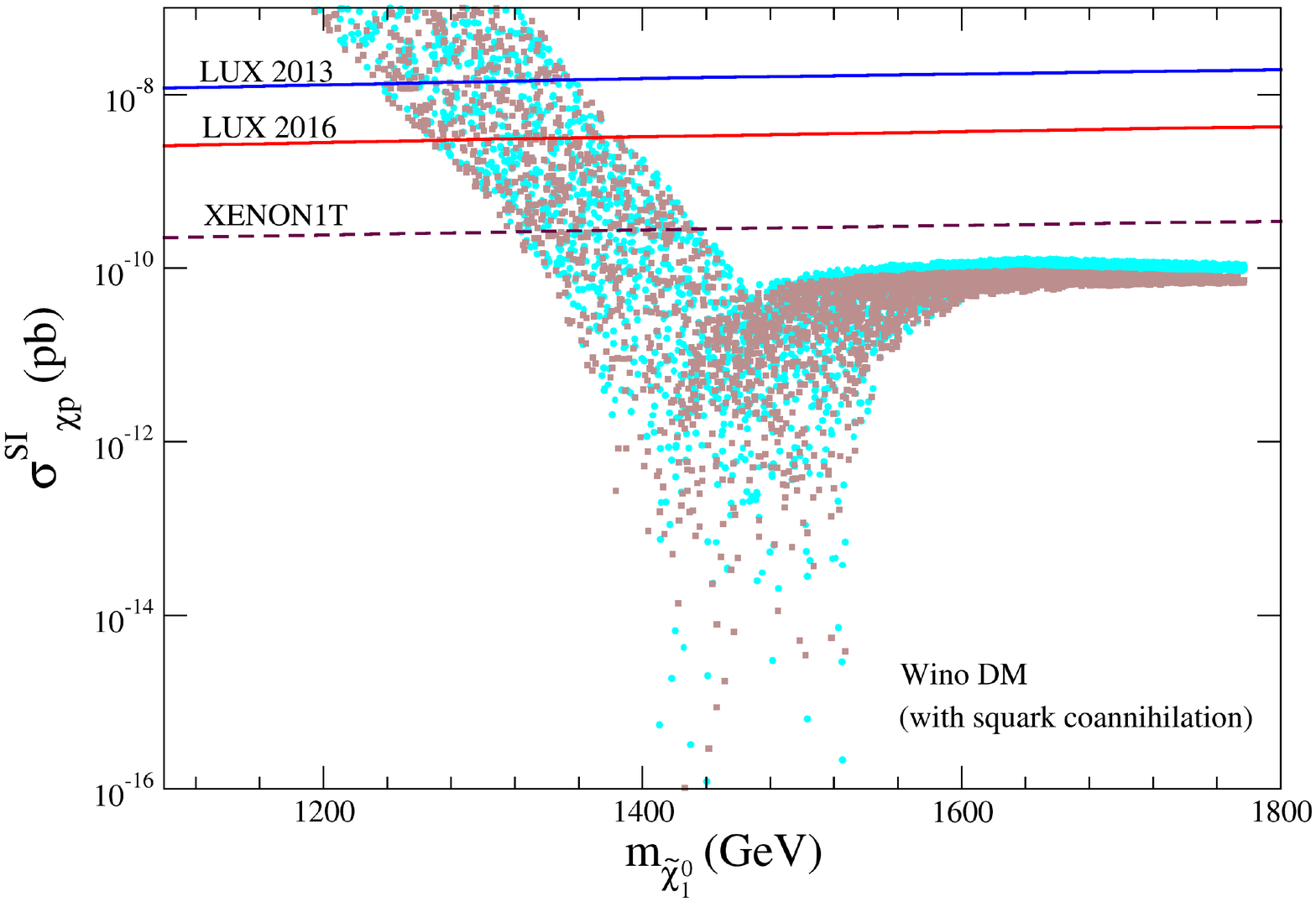}\\
   \begin{center}
  \mygraphthree{wino_si_slep_sq}{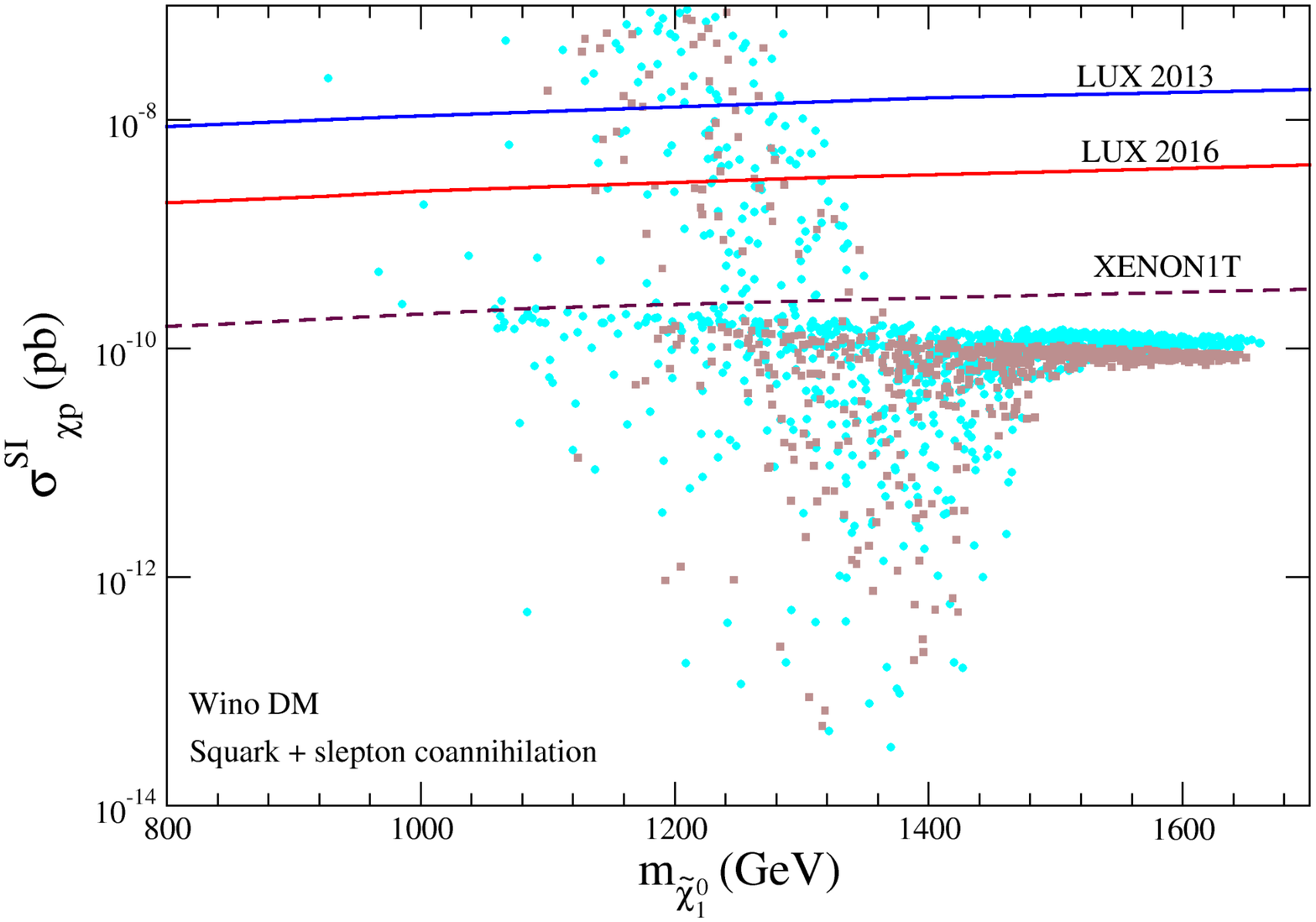}
  \end{center}
  \caption{\small (a) Scatter plot of spin-independent DM 
direct detection cross-section
    vs LSP mass 
      for wino dominated LSPs undergoing   
      LSP-slepton coannihilations. 
The cyan and brown points represent tan$\beta=$ 10 and 30 respectively 
 that satisfy Eq.(\ref{planckdata}). The red line (solid) is the LUX 2016
      exclusion 
contour and the maroon dashed line shows the expected limit from the future
XENON1T experiment. (b) Same as (a) except LSP undergoing LSP-squark 
coannihilations. (c) Same as (a) except LSP undergoing slepton plus squark 
coannihilations.
}
  \label{wsi_slep_sq_caption}
\end{figure}

From the above expressions it is clear that 
the couplings and hence the scattering cross-section would be large 
if there is a large degree of mixing between the gaugino and 
the higgsino components of the LSP. 
We also note that couplings become weaker 
for increased gaugino masses and $\mu$.  
On the other hand, a pure higgsino or a wino LSP 
with very little mixing 
can hardly be able to produce large values of spin-independent cross-section. 
Fig.\ref{hsi_slep_sq_caption} shows our results for DM direct SI 
detection cross-section where only
the points satisfying Eq.\ref{planckdata} are shown for a  
higgsino type of LSP undergoing LSP-slepton coannihilations 
(Fig.\ref{higgsino_si_slep}) and LSP-squark coannihilations (Fig.\ref{higgsino_si_sq}).  
The cyan and brown points correspond to tan$\beta=$ 10 and 30 
respectively.  
The red line (solid) is the LUX 2016 exclusion 
contour\cite{luxnew} and the 
maroon dashed line shows the expected limit from the future
XENON1T experiment\cite{xenon1t}.
Clearly, the recent LUX data rules out low higgsino mass 
region below 600~GeV (Fig.\ref{higgsino_si_slep}). 
We must also remember the existence 
of uncertainty, around one order in magnitude, 
in the computation of the SI direct detection 
cross-section. Factors like strangeness content of nucleon, 
local DM density, velocity distribution profiles, all contribute toward 
such uncertainty amount (See Ref.\cite{ourwork1} and references therein). 
As seen in Fig.\ref{higgsino_si_slep} the higgsino LSP scenario may be 
effectively probed via XENON1T.  
Fig.\ref{higgsino_si_sq} shows similar results for 
the LSP-squark coannihilations. Here, the lowest LSP mass
that survives after the LUX 2016 data is about 840~GeV. 
Additionally, a large region 
of parameter space\footnote{By parameter space one really means here a
smeared region of squark masses around a given LSP mass.} 
is discarded via the same experiment without however 
affecting the lowest possible value of the LSP mass.    
Compared to the case of Fig.\ref{higgsino_si_slep} here the
 SI direct detection cross-sections are generally large.
 This is a signature of having a quasi-degenerate squark and LSP scenario that 
comes into our study of the LSP-squark coannihilations. Here
the effective coupling constant for quark-LSP scattering
drastically increases\cite{wells_murakami} 
causing the cross-section to be larger, often much more
than the LUX limit for a significant zone of the LSP mass. 
  Thus, the squark exchange diagrams are potentially able to 
  compete with or even dominate over the Higgs exchange diagrams
  while contributing to
  the total direct detection cross-section\footnote{The extent of 
  any cancellation effect on the contrary is small, unlike the wino
  scenario that we will see soon.}. This is of course true
  for some region of parameter 
  space where the degree of the LSP-squark mass degeneracy is higher.  
  We remind that the above is unlike the usually
  encountered MSSM parameter regions where 
  Higgs exchange diagrams dominate over the 
  squark exchange diagrams in the SI direct detection cross-section.
  Coming back to Fig.\ref{higgsino_si_sq} we see that a lot of parameter
  space is eliminated via LUX 2016 data. The remaining parameter space  
  can fully be probed in the XENON1T experiment. The effect of including
  both slepton and squark coannihilations is shown in
  Fig.\ref{higgsino_si_slep_sq}. 
  The lowest LSP mass limit satisfying the LUX data is around 680~GeV.

Fig.\ref{wsi_slep_sq_caption} shows our results for DM direct SI 
detection cross-section where only
the points satisfying Eq.\ref{planckdata} are shown for a  
wino type of LSP undergoing slepton coannihilations 
(Fig.\ref{wino_si_slep}) and squark coannihilations 
(Fig.\ref{wino_si_sq}).  
The cyan and brown points correspond to tan$\beta=$ 10 and 30 
respectively.  
The red line (solid) is the LUX 2016 exclusion 
contour\cite{luxnew} and the 
maroon dashed line shows the expected limit from the future
XENON1T experiment\cite{xenon1t}.
Clearly, as seen in Fig.\ref{wino_si_slep} 
the cross-section is too low so that even XENON1T would 
not be able to probe this scenario except around the 1.1~TeV region for 
LSP mass. We must additionally clarify that 
a comparison of Fig.\ref{higgsino_si_slep} and Fig.\ref{wino_si_slep} 
shows that contrary to what we would naively expect,
the SI cross-section in the latter 
case is in general smaller. The reason lies in the 
fact that the values of $\mlspone$ that satisfy the relic density
limits for a wino like LSP are much higher than that of a higgsino dominated 
LSP.
Fig.\ref{wino_si_sq} shows similar results for 
the LSP-squark coannihilations. Here, the lowest LSP mass
that survives after the LUX 2016 data is about 1.27~TeV. 
Additionally, a large region 
of parameter space is discarded via the same experiment while  
eliminating a window of 1.1~TeV to 1.27~TeV of LSP mass. XENON1T would not be 
able to probe this scenario except the region close to 1.2 TeV of the LSP 
mass.
Similar to the case of higgsino-squark coannihilations, 
the SI direct detection cross-section
is much larger for wino-squark coannihilation scenario compared
to the wino-slepton results.
The squark exchange contributions can be significantly
large for the parameter points associated with near degeneracy of
the squark and LSP masses as explained before.  
Here, we observe that the cross-section can be
quite large even for squark masses $\gsim$~1.2~TeV. The Higgs-exchange
contributions are sub-dominant in this case of wino-dominated LSP 
that has a very small higgsino content. 
We must also note that for a fixed value of $\mlspone$ we
get a wide degree of variation in cross-section with some points  
exceeding the LUX
 limit whereas some points having values below the limit.  In the latter case
there is a cancellation among the contributions from the squark 
and Higgs exchange diagrams that pushes the cross-section to very low
values. Similar to what happens for the higgsino LSP case as
in Fig.\ref{higgsino_si_sq}, this
is a signature of quasi-degenerate squarks and LSP that
leads to a large increase in the effective coupling constant for quark-LSP
scattering\cite{wells_murakami}.

Finally, the effect of including
both slepton and squark coannihilations is shown in Fig.\ref{wino_si_slep_sq}. 
The lowest LSP mass limit satisfying the LUX data is about 1~TeV.

We will come to the discussion of the notch regions of Fig.\ref{higgsino_delta_sq} and Fig.\ref{wino_delta_sq}. 
Apart from direct detection, enhancement of $q{\tilde q}{\tilde \chi}$
effective coupling near the 
degenerate zone of squarks and LSP masses has its important
signature also on the DM relic density. 
For a wino dominated LSP that does not have a quasi degenerate 
neutralino state, the notch region 
is found to coincide with the mentioned cancellation region of
$\sigma^{SI}_{\chi p}$ (i.e. cancellation between the higgs exchange and the
squark exchange diagrams). Corresponding to a given mass of the
LSP, this is the region of $\delta$ where the above effective
coupling becomes large.
The situation for a higgsino-LSP case is more involved.
This is principally because on the top of the coannihilations a
wino LSP would undergo,
there are  additional coannihilation processes like
$\lsptwo \lspone$, $\lsptwo {\tilde q}$ contributing
toward the higgsino relic density.
As a result, in spite of a 
cancellation zone of $\sigma^{SI}_{\chi p}$
for certain values of $\delta$, coannhilation effects 
potentially smear the abrupt change in the higgsino relic density
coming out of the effect of enhanced
$\lspone {\tilde q} q$ and $\lsptwo {\tilde q} q$
coupling strengths. Consequently, for a higgsino DM
the values of $\delta$ that correspond to a cancellation or
an enhancement zone in $\sigma^{SI}_{\chi p}$ are not the 
same where the anomalous ``notch'' zone of the relic density occurs. However,
the enhancement of coupling remains a valid fact. It is
seen that for a given $\mlspone$ there is an abrupt decrease of the
DM relic density corresponding to some range of $\delta$. Once
a lower and a upper limit of the relic density are
imposed, the above decrease in relic density irrespective of the
LSP mass, leads to the formation of the notch regions for some effective
range of values of $\delta$.
Details may be explained by examining the relevant coupling enhancements   
as given in Ref.\cite{wells_murakami}\footnote{Specifically we refer Eqs. 15 to 17 and A8 to A14 of the paper.}.

\subsection{Indirect Detection}
\label{indirectDMsub}
DM particles may get trapped due to gravity inside astrophysically dense
objects like the Sun or the Earth 
by losing energy through repeated scattering 
with the nucleons. Inside the core of these objects 
DM particles may undergo pair annihilations
leading to SM particles like fermion-antifermion pairs, 
gauge bosons etc. in the final state. 
The resulting antiparticles, neutrinos and gamma rays 
can offer interesting indirect signals of DM in the galaxy.

The high energy neutrinos produced as end products of 
DM pair annihilation in the solar core 
can produce muons through charged current interactions.
The IceCube experiment\cite{icecube1} provides bounds on the muon flux 
for the pair annihilation channel DM DM $\rightarrow W^+ W^-$.
Fig.\ref{hmufluxslep} shows a scatter plot of the values of muon flux 
as a function of $\mlspone$ for higgsino dominated LSPs undergoing 
slepton coannihilations for parameter points satisfying Eq.\ref{planckdata}. 
The cyan and brown points correspond to $\tan\beta=10$ and 30 respectively.
The green and blue lines are the current and projected
limits from the IceCube experiment\cite{icecube1,icecube2}
in the $\lspone \lspone \rightarrow W^{+} W^{-}$ channel. 
Similarly, Fig.\ref{hmufluxsq} shows the results for LSP-squark 
coannihilations. Clearly, the fluxes in both the above figures are 
in general too small to be probed. 

\begin{figure}[!htb]
\vspace*{0.3in}
\vspace*{-0.05in}
\mygraph{hmufluxslep}{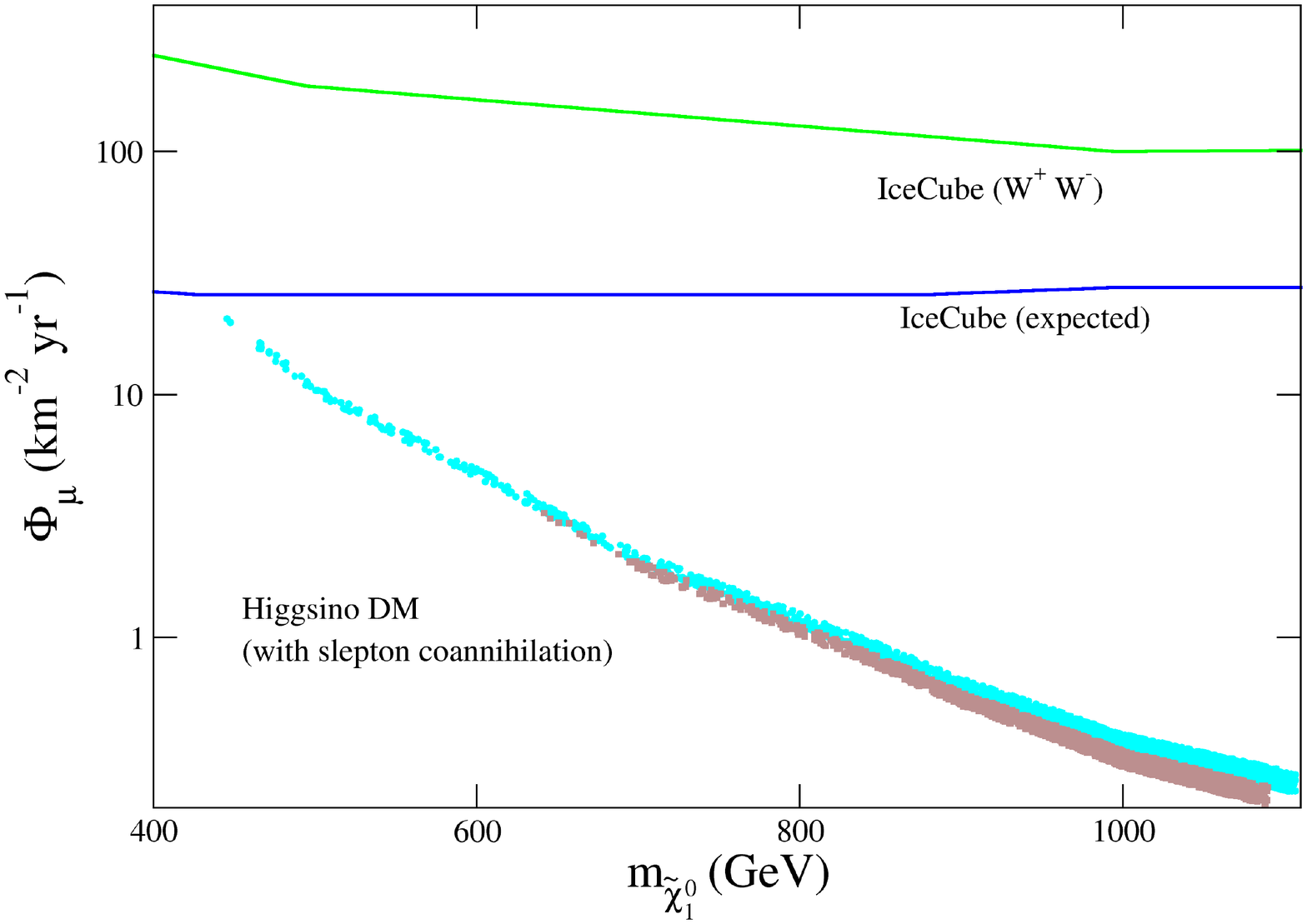}
\hspace*{0.5in}
\mygraph{hmufluxsq}{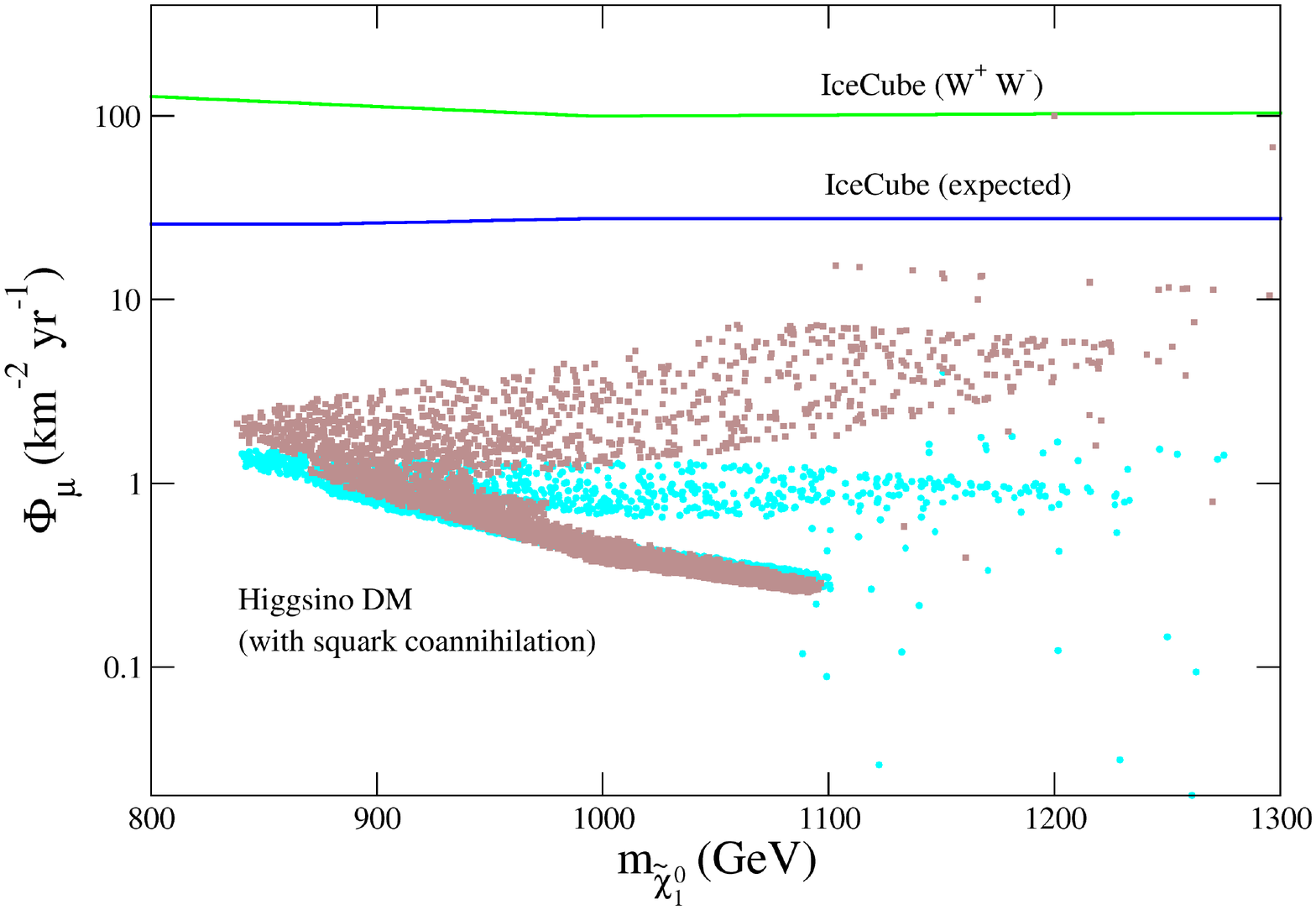}
\caption{\small (a) Scatter plot of muon flux 
vs $\mlspone$ for higgsino LSPs undergoing 
LSP-slepton coannihilations for $\tan\beta=10$ (cyan) and 30 (brown) while 
satisfying Eq.(\ref{planckdata}). Present and future IceCube limits 
are shown as green and blue lines respectively. 
(b) same as (a) except the LSP is undergoing LSP-squark coannihilations. 
}
\label{hmuf_slep_sq_caption}
\end{figure}

\begin{figure}[!htb]
\vspace*{0.3in}
\vspace*{-0.05in}
\mygraph{wmufluxslep}{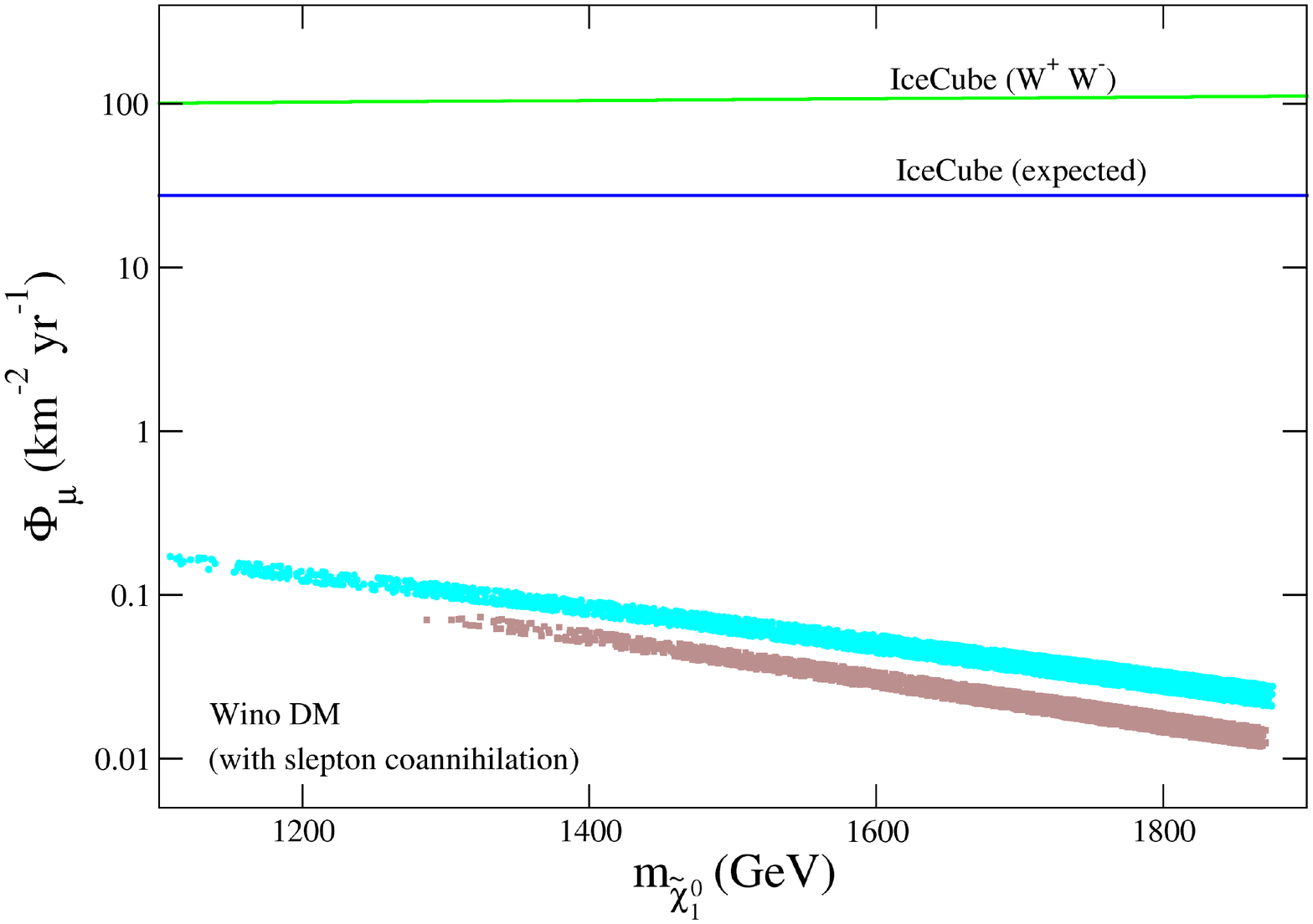}
\hspace*{0.5in}
\mygraph{wmufluxsq}{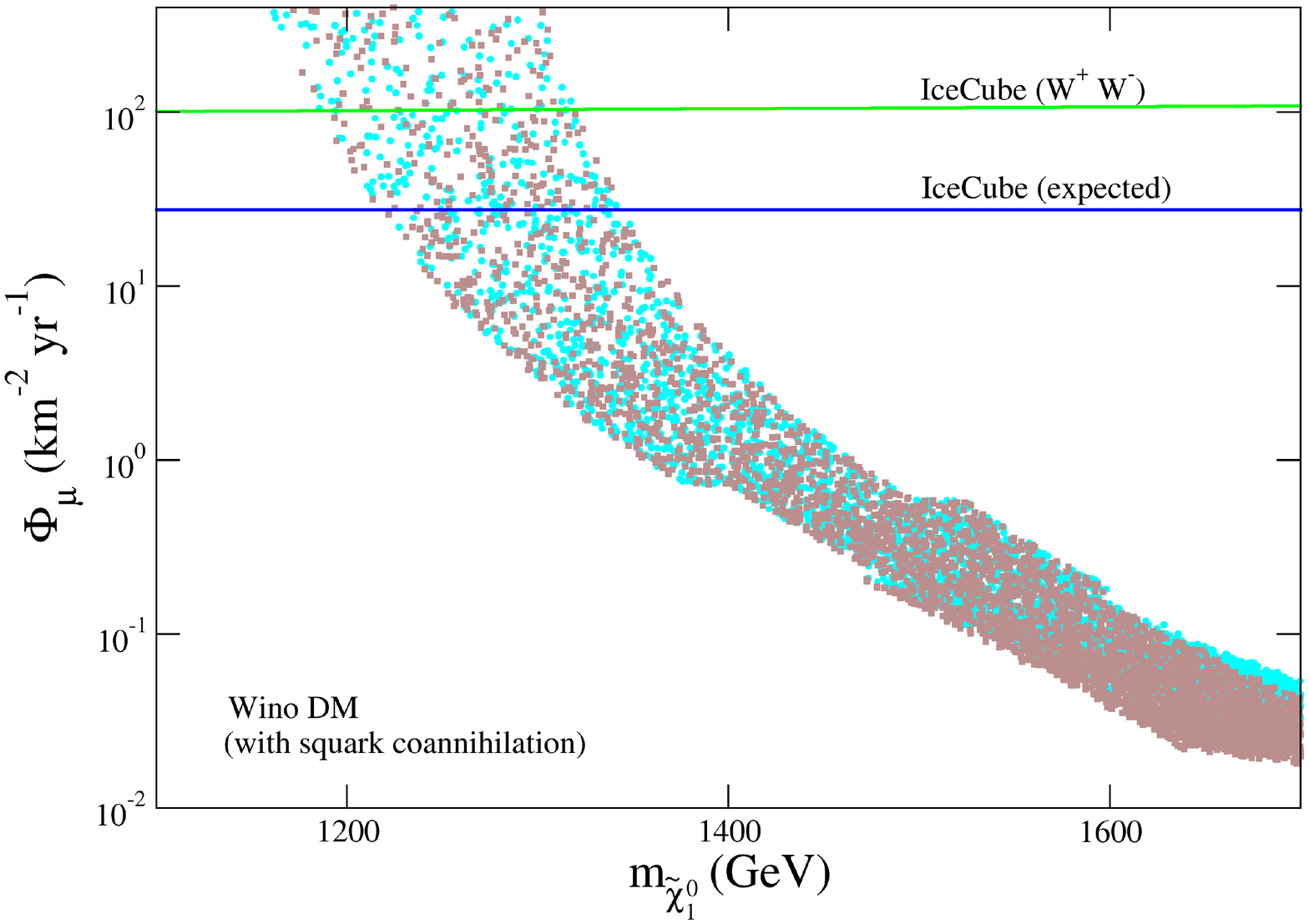}
\caption{\small (a) Scatter plot of muon flux 
vs $\mlspone$ for wino LSPs undergoing 
LSP-slepton coannihilations for $\tan\beta=10$ (cyan) and 30 (brown) while 
satisfying Eq.(\ref{planckdata}). Present and future IceCube limits 
are shown as green and blue lines respectively. 
(b) same as (a) except the LSP is undergoing LSP-squark coannihilations. 
}
\label{wmuf_slep_sq_caption}
\end{figure}

The results of muon flux for the case of a wino dominated LSP is 
shown in Fig.\ref{wmuf_slep_sq_caption}. 
Fig.\ref{wmufluxslep} shows the results for LSP-slepton coannihilations 
for parameter points 
that satisfy Eq.\ref{planckdata} for wino dominated LSPs undergoing 
slepton coannihilations.
The color convention and the details of the limits from IceCube data 
are similar to Fig.\ref{hmufluxslep}.
Clearly, the flux is too small to be probed for the LSP-slepton coannihilation 
scenario.
Similarly, Fig.\ref{wmufluxsq} shows the results for LSP-squark 
coannihilations. The result does not show any more exclusion of low mass LSP 
region compared to what is seen in Fig.\ref{wino_si_sq} for 
the SI direct detection cross section.  

Let us now discuss the constraints on spin-dependent 
DM-nucleon interaction cross-section as derived from the IceCube data. 
Inside the solar core, the number density $N$ of DM particles at any 
instant of time $t$ is obtained from the following\cite{Kamionkowski},
\begin{equation}
\frac{dN}{dt} = C_c - C_A N^2, 
\label{indirect}
\end{equation}
where, $C_c$ is the capture rate of DM by interaction 
with the nucleons present at the surface of the Sun whereas $C_A$ 
is related to the annihilation rate $\Gamma_A$ as : 
$\Gamma_A = \frac{1}{2} C_A N^2$.  Solution of Eq.\ref{indirect} leads to
$\Gamma_A =\frac {1}{2} C_c$ tan$h^2(t/\tau)$, 
with $\tau = \frac {1}{\sqrt{C_c C_A}}$.  Hence, the capture rate is determined
by the annihilation rate and when the age of the universe is 
much greater than $\tau$ (which occurs for large 
$C_c$ and $C_A$), an equilibrium is reached so that  
$\Gamma_A = \frac {1}{2} C_c$.   Thus, it is possible
to put bounds on the annihilation and capture cross-sections 
by looking at the indirect DM signals from the Sun.
Since, capture of the DM particles occurs through 
spin-independent/dependent (SI/SD) DM interactions with the nucleons, 
these bounds get translated into the bounds on DM SI/SD interaction 
cross-sections.

$\lspone$ can have spin-dependent interaction with the 
quarks via s-channel squark exchange and t-channel $Z$-boson
exchange processes. Similar to the SI case, while considering
LSP-slepton coannihilations, we can safely
ignore the contributions from the squark exchange processes
since the squarks are taken to be heavy.  
The tree level $Z \lspone \lspone$ coupling is given 
by $c_{Z \lspone \lspone} = \left(N_{13}^2 - N_{14}^2\right)$.
For the higgsino LSP case the coupling is given as\cite{hisano_nojiri},
\begin{eqnarray}
c_{Z \lspone \lspone}
&\simeq&
\mp \frac12 \left(t_W^2 \frac{m_W^2}{M_1\mu}+\frac{m_W^2}{M_2\mu}\right)
\cos2\beta +O\left(\frac{\mu}{M_1},\frac{\mu}{M_2}\right),
\label{higgsino_z}
\end{eqnarray}
with $\mu>0$$(\mu<0)$. 
The same coupling for the wino case takes the form\cite{hisano_nojiri},
\begin{eqnarray}
c_{Z \lspone \lspone}
&\simeq&
\frac{m_W^2}{M_2^2-\mu^2} \cos2\beta.
\label{wino_z}
\end{eqnarray}
Thus, in general the couplings get suppressed as the LSP,
irrespective of a higgsino or a wino becomes heavy.  
Fig.\ref{hsdslep} shows the results for the SD cross section 
for the higgsino dominated LSP scenario for LSP-slepton coannihilation. 
Fig.\ref{hsdsq} shows the
LSP-squark coannihilation case for which the degeneracy between squark and
the LSP masses (similar to what was described in the SI case)
may push up the SD cross section.
In general the IceCube limits would be inadequate to probe
such higgsino models. 

\begin{figure}[!htb]
\vspace*{0.3in}
\vspace*{-0.05in}
\mygraph{hsdslep}{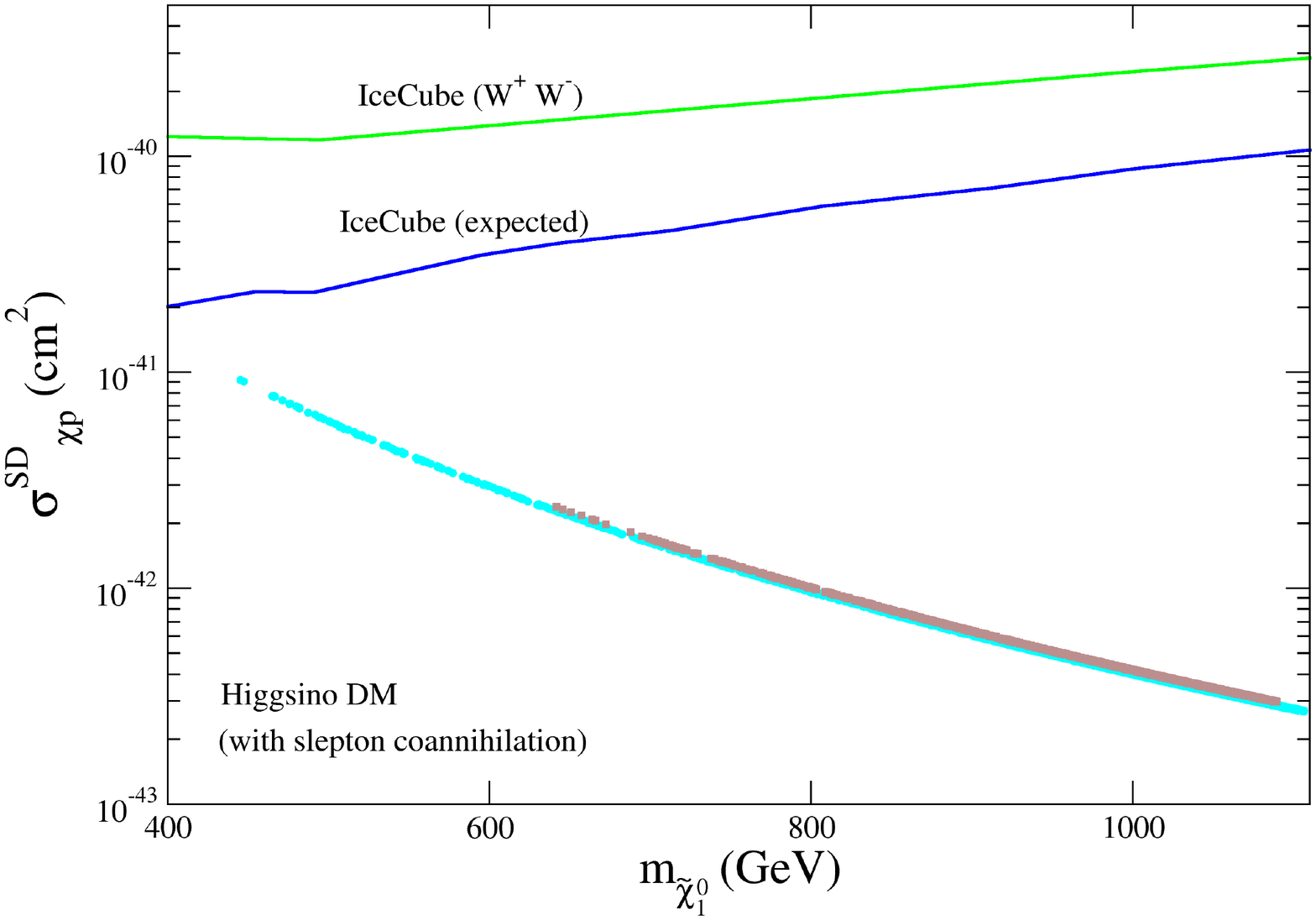}
\hspace*{0.5in}
\mygraph{hsdsq}{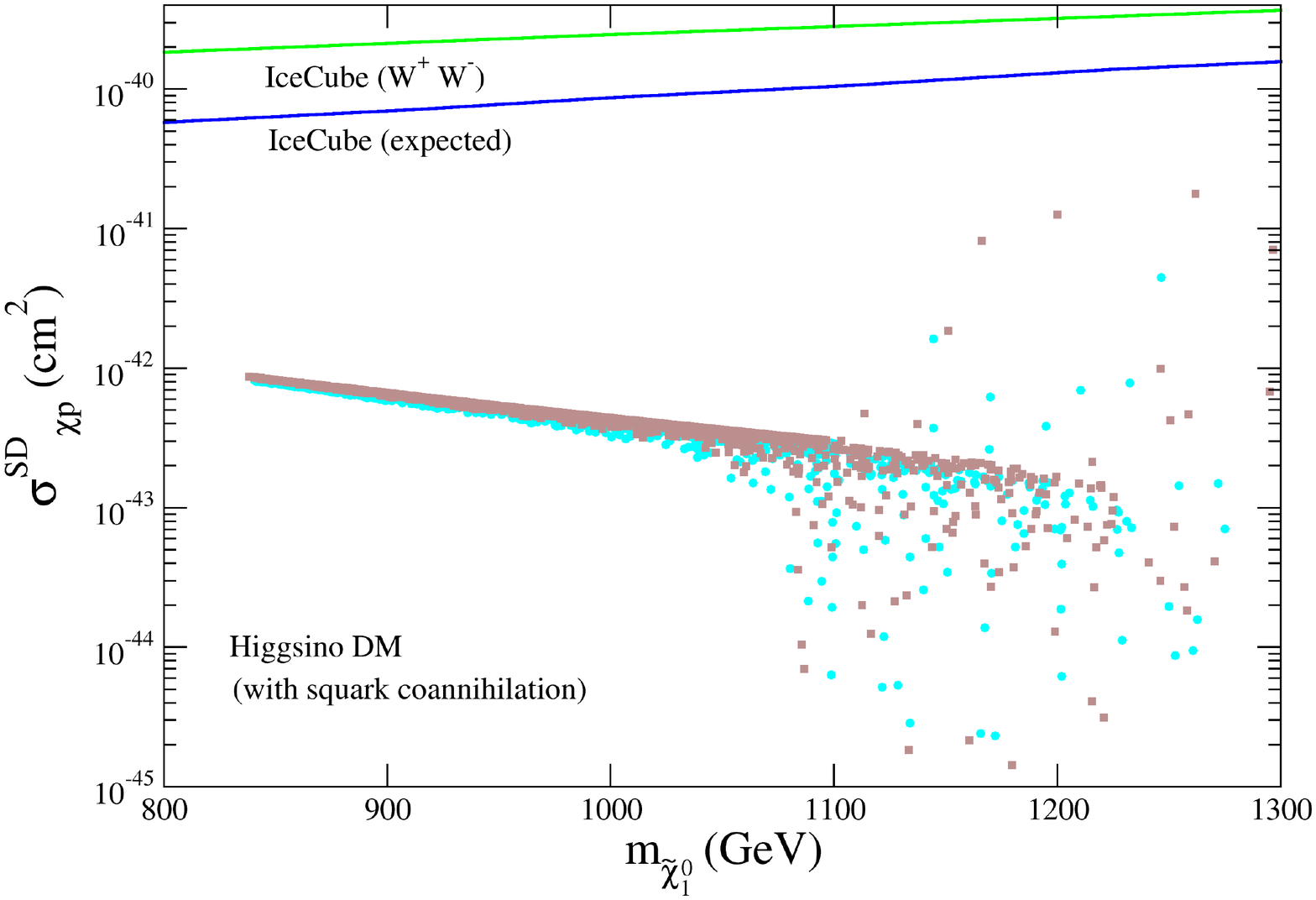}
\caption{\small (a) Scatter plot of  spin-dependent 
direct detection cross-section vs $\mlspone$ for higgsino LSPs undergoing 
LSP-slepton coannihilations for $\tan\beta=10$ (cyan) and 30 (brown)
while satisfying Eq.(\ref{planckdata}). 
Present and future IceCube limits are shown as 
green and blue lines respectively.  (b) Same as (a) except the 
LSP is undergoing LSP-squark coannihilations.}
\label{hsd_slep_sq_caption}
\end{figure}

\begin{figure}[!htb]
\vspace*{0.3in}
\vspace*{-0.05in}
\mygraph{wsdslep}{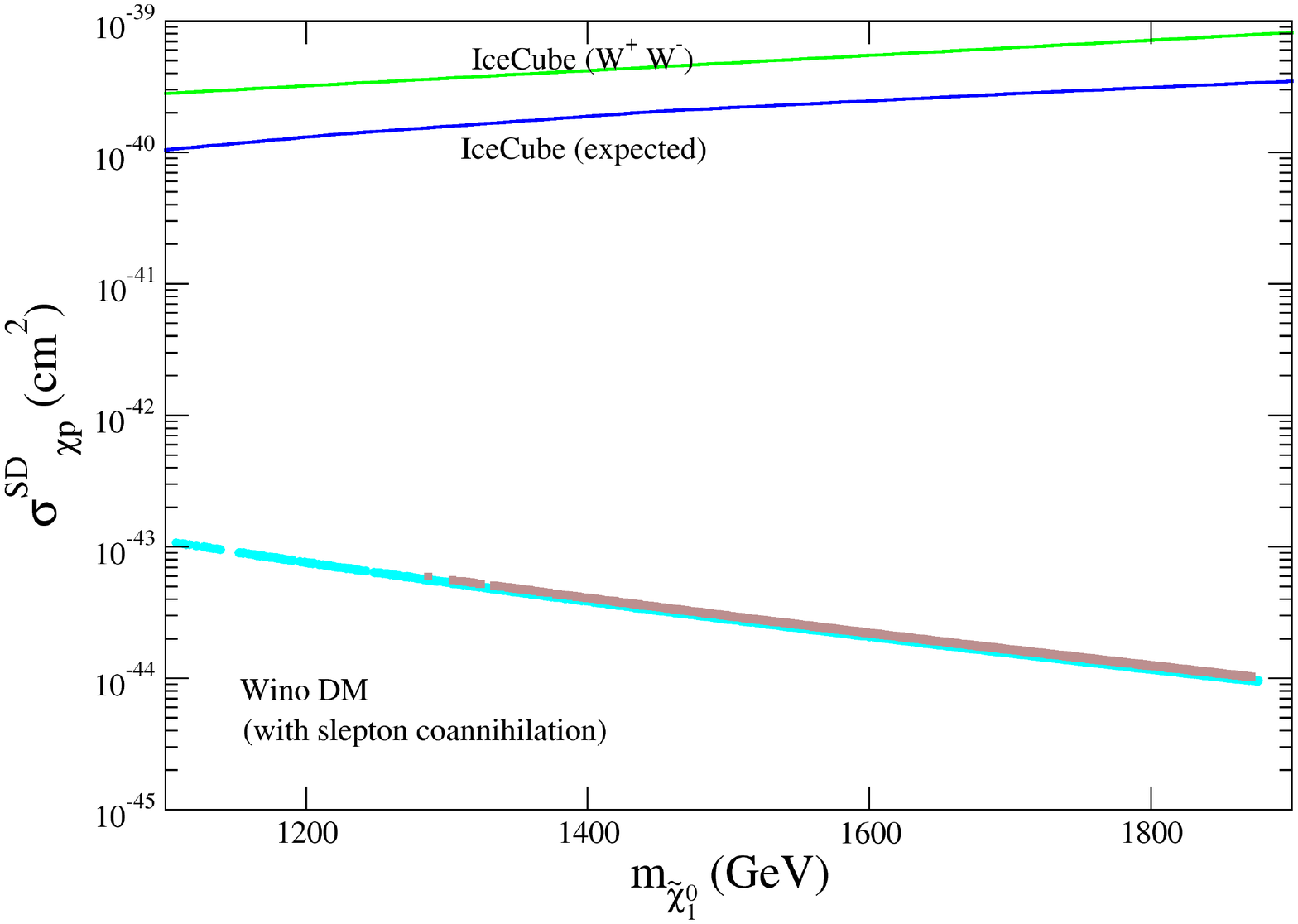}
\hspace*{0.5in}
\mygraph{wsdsq}{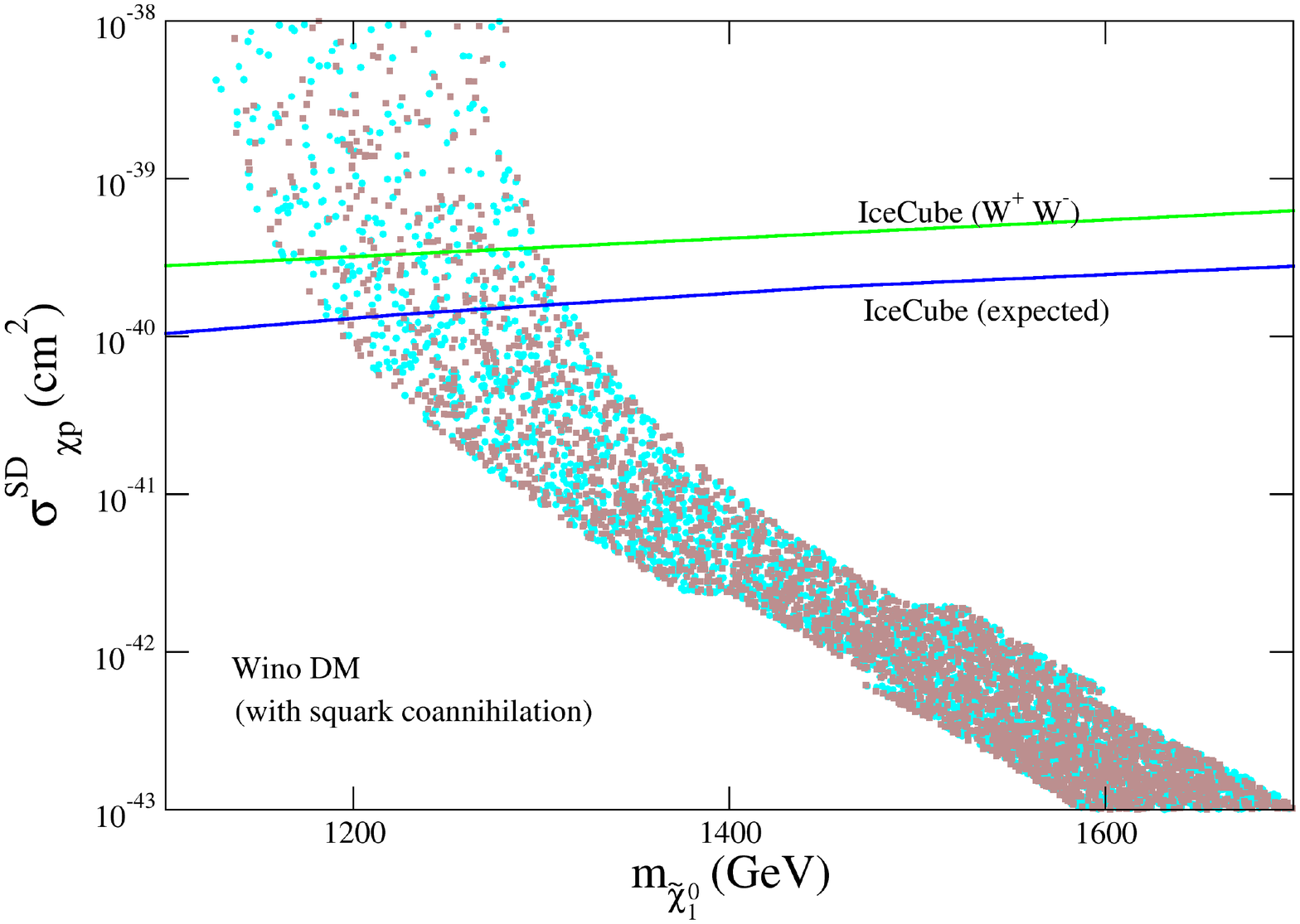}
\caption{\small (a) Scatter plot of  spin-dependent 
direct detection cross-section vs $\mlspone$ for wino LSPs undergoing 
LSP-slepton coannihilations for $\tan\beta=10$ (cyan) and 30 (brown) while 
satisfying Eq.(\ref{planckdata}).
Present and future IceCube limits are shown as 
green and blue lines respectively.  (b) Same as (a) except the 
LSP is undergoing LSP-squark coannihilations.}
\label{wsd_slep_sq_caption}
\end{figure}

Fig.\ref{wsd_slep_sq_caption} shows the results for the SD cross section 
for the wino dominated LSP scenario for LSP-slepton coannihilation 
(Fig.\ref{wsdslep}) and LSP-squark coannihilation (Fig.\ref{wsdsq}) cases. 
Although the IceCube limits may eliminate some region of 
parameter space where the LSP undergoes squark coannihilations,    
the result does not show any more exclusion of low mass LSP 
region compared to what is seen in Fig.\ref{wino_si_sq} for 
the SI direct detection cross section.  

In Table~\ref{bptable} we show two benchmark points (BP)
satisfying WMAP/PLANCK relic density limits of Eq.(\ref{planckdata})
as well as the direct and indirect detection limits from the LUX
and IceCube experiments respectively.
BP1 and BP2 correspond 
to the case of a higgsino-LSP undergoing
slepton and squark coannihilations for masses
617~GeV and 760~GeV respectively for $\tan\beta =10$. 
Monojet searches at the 14 TeV LHC can probe pure higgsino scenario 
only upto the mass of $\mlspone \sim $ 410 GeV\cite{low_wang2014}.  
However, the situation looks more promising for a 100 TeV collider 
where higgsinos may be probed upto 1.2 TeV. 
Existing disappearing track searches at the LHC
do not have much sensitivity to a higgsino LSP. However, with modifications in 
search strategy, as suggested in Refs.\cite{higgsinoDTnew2017}, 
higgsinos upto $\sim$ 600 GeV and $\sim$ 1.1 TeV could
be probed by the 14 TeV high luminosity (HL)-LHC and 
a 100 TeV collider respectively. 
BP3 and BP4 refer to wino-like LSP participating in 
slepton and squark coannihilations with masses 1011~GeV and 1188~GeV
respectively for the same value of $\tan\beta$. Although a HL-LHC at 14 TeV
seems to be unable to probe these benchmark points, a 100 TeV collider, with 
an exclusion reach of $\sim$ 1.8 TeV in the monojet search channel
can decisively explore such scenarios\cite{low_wang2014}. 
These benchmark scenarios for 
wino-like DM is likely to evade the HL-LHC even with disappearing track 
searches. However, the same 
searches at a 100 TeV collider can conclusively probe these cases.
Apart from collider searches, all the four BPs will be probed in near future
with the XENON1T experiment. However, they are unlikely to produce any signal
in future indirect detection experiments.

\clearpage
\begin{table}[H]
{\small
\begin{center}
\begin{tabular}{|c|c|c||c|c|}
\hline
Parameters & \multicolumn{2}{c|}{Higgsino DM} & \multicolumn{2}{c|}{Wino DM} \\
\cline{2-5}
\hline
Points & BP1    & BP2  & BP3  & BP4\\
\hline
$M_1$ (GeV) & 1221.5    & 1502.8  & 2339.2  & 2847.8\\
\hline
$M_2$ (GeV) & 1465.8    & 1803.4 & 974.7 & 1186.6\\
\hline
$\mu$ (GeV) & 610.8  & 751.4 & 1949.3   & 2373.2\\
\hline
$M_{\tilde q_{L,R}}$ (GeV) & 3000 & 539.6 & 4000 & 944.4\\
\hline
$M_{\tilde l_{L,R}}$ (GeV) & 628.4 & 3000 & 1026.3 & 4000\\
\hline
$\mlspone$ (GeV) & 617.3  & 760.5  & 1011.1 & 1188.5\\
\hline
$\mchonepm$ (GeV) & 620.4  & 763.1 & 1011.1 & 1188.5\\
\hline
$\mlsptwo$ (GeV) & 1499  & 767  & 1969 & 2392\\
\hline
$\mlspthree$ (GeV) & 1198  & 1471.8  & 1969.8 & 2393 \\
\hline
$M_{\tilde (e,\mu)_{L}}$ (GeV) & 630 & 3000 & 1027.2 & 4000\\
\hline
$M_{\tilde (e,\mu)_{R}}$ (GeV) & 630   & 3000  & 1027.2 & 4000\\
\hline
$\mstauone$ (GeV) & 621.7  & 2998  & 1011.1 & 3995.2\\
\hline
$\mstautwo$ (GeV) & 638  & 3000  & 1043 & 4005.2\\
\hline
$M_{\tilde \nu}$ (GeV) & 625.3  & 2999  & 1024.4 & 3999.5\\
\hline
$M_{\tilde u_L}$ (GeV) &  3000  &  792.8 & 4130.7 & 1228 \\ 
\hline
$M_{\tilde d_L}$ (GeV) &  3000  &  796.8 & 4131.4 & 1230.5 \\
\hline
$M_{\tilde u_R}$ (GeV) &  3000  &  793.7 & 4130.9 & 1228.5 \\
\hline
$M_{\tilde d_R}$ (GeV) &  3000  &  795   & 4131.1 & 1229.4 \\
\hline
$M_{{\tilde t}_1}$ & 2905.5 & 2889.0 & 3956.1 & 3962.6 \\
\hline
$m_h$ & 126.5  & 126.4 & 126.5 & 125.9 \\
\hline
$\Omega_{\tilde \chi} h^2$ &  0.126 &  0.092  & 0.091 & 0.09\\
\hline
$\sigma_{SI} $ $\times 10^{-9}$ (pb)& 1.39  & 1.69  & 0.33 & 1.21\\
\hline
$\sigma_{SD} $ $\times 10^{-6}$ (pb)& 2.64  & 1.2 & 0.15 & 11.3\\
\hline
$\Phi_\mu  (km^{-2} yr^{-1}) $ &  4    & 2.1  & 0.2  & 4 \\
\hline
\end{tabular}
  \end{center}
    \caption{ Table of benchmark points allowed by WMAP/PLANCK data of 
Eq.(\ref{planckdata}) as well
as direct detection bounds from the LUX and indirect detection constraints
from the IceCube. All the benchmark points are for $\tan\beta=10$.
BP1 (BP3) and BP2 (BP4)
correspond to the case of a higgsino (wino)-LSP undergoing slepton and squark
coannihilations respectively. The relevant SM parameters used are $m_t^{pole}=173.2$~GeV, $m_b^{\overline{MS}}=4.19$~GeV and $m_\tau=1.77$~GeV.}
\label{bptable}
}
\end{table}


\section{Conclusion}
\label{conclusion}
A bino-dominated LSP generally produces overabundant DM. A bino-like LSP 
relies mostly on the 
bulk-annihilation or t-channel slepton exchange mechanism, a disfavored
scenario in the context of LHC data.
A bino can also be a DM candidate with the help of
coannihilations with sleptons (typically staus), or coannihilations with
suitable electroweakinos as in the Focus Point/Hyperbolic Branch
region, or it can take the help of
s-channel Higgs mediation for pair annihilation in order to satisfy the
DM relic density limits. 
On the other hand, in MSSM there are theoretical as well as
phenomenological motivations to study higgsino and wino-dominated LSPs.
When the LSP turns out to be a higgsino, 
these processes
include pair-annihilation and coannihilations among $\lspone, \chonepm$
and $\lsptwo$. The same for a wino-LSP situation include
coannihilations between $\lspone$ and $\chonepm$.  
It is known 
that these processes are too strong to cause the LSP to become
an underabundant component of DM 
unless its mass is around 1~TeV for a higgsino or a little above 2~TeV for 
a wino type of LSP. 
We consider a compressed scenario of pMSSM where sfermions may 
take a very significant role as coannihilating sparticles. 
Our purpose is to examine 
how light the LSP as a higgsino or a wino can be while it satisfies 
both the lower and the upper limit  
of the DM relic density as given by the WMAP/PLANCK data.
We choose
two representative values of $\tan\beta$, namely 10 and 30 and
consider both sleptons and squarks as coannihilating partners.
In regard to the 
LSP-slepton coannihilations we consider all the three generations of 
sleptons including also the sneutrinos while keeping the squark masses
heavy. We perform the analysis 
by requiring a maximum of 20\% mass difference between that of
the LSP and each of 
its coannihilating partners. Consideration of the slepton
coannihilations reduces the effective cross section leading to an increase 
in the relic density.  This is how the relic density gets modified or 
in other words this is how the lower limit of the LSP mass satisfying the 
relic density limits decreases. We find that for a higgsino dominated LSP
the lowest LSP mass that satisfies the relic density limits is about
450 GeV, about a 60\% reduction corresponding to the case of
no sfermion coannihilations and this occurs for $\tan\beta=10$.
The dependence on $\tan\beta$ comes via
  the L-R mixing of tau-sleptons and the exponential suppression
  generically associated with coannihilation toward the effective annihilation
  cross-section.
The same reduction in the lower limit
for a wino type of LSP occurs for $\tan\beta=10$ and it is about 1.1~TeV, 
more than a 100 percent reduction in the value corresponding to the case of
no sfermion coannihilations.   
For squarks, we allowed coannihilations with  
only the first two 
generations of squarks while imposing a similar 20\% limit as before for the
deviation of masses of the coannihilating particles from the LSP mass
keeping the
third generation of squarks as well as sleptons of all the generations very heavy. 
The reason of omitting the third generation lies in the fact that
a large splitting between the two top-squark masses as required by a
125 GeV Higgs boson would take away a lot of parameter space if 
  we need a uniform 20\% limit for the difference of each of the 
  squarks and the LSP masses.
In the absence of coannihilating third generation of squarks,
our results become essentially independent of $\tan\beta$.
The lowest LSP mass satisfying the relic density limits is about 840~GeV
for the higgsino case, only a modest reduction by 10-15\%
from the generic higgsino LSP scenario. For 
  the higgsino-squark coannihilation scenario we additionally obtain a
  region of
  parameter space where the relic density is decreased when squark
  coannihilations come to the picture, thus increasing the
  upper limit of the LSP mass satisfying the relic density data. This happens
  only in a very limited zone of parameter space with
  nearly degenerate squark and LSP masses and 
  toward the end of the upper limit of the LSP mass 
  satisfying the DM relic density data. 
Coming to wino, the lower limit of the LSP
mass with the above squark coannihilations
is around 1.1~TeV. Additionally, computation for a scenario of 
combined slepton and squark coannihilations shows that the lower limit of
higgsino-LSP is about 500~GeV whereas for a wino-LSP the same
is about 900~GeV. We also note that throughout our study
  we consider the CP-odd Higgs boson ($A$) to be 
sufficiently heavy so as to avoid an s-channel $A$-pole.

We further analyze the direct and indirect detection prospects of DM for the
above types of LSPs for the two kinds of sfermion coannihilations
considered in this work. 
In the part of the analysis that involves squark-LSP coannihilations, 
  because of the near degeneracy of squarks with the LSP,
  the squark exchange diagrams in the direct detection cross section
  can be very important. These may even exceed the contributions   
  from the Higgs exchange diagrams which typically dominate the generic MSSM
  parameter space. 

The SI direct detection cross section may
exceed the recent LUX data for a higgsino type of LSP undergoing slepton coannihilations for a mass below 600~GeV. For squark
coannihilations, the above number is about 840~GeV.
The corresponding number for the case of slepton plus squark coannihilations
is around 680~GeV.
The same occurs at around 1.27~TeV for
 a wino-LSP undergoing squark coannihilations whereas there is no direct detection
 constraint for the part of the study involving slepton coannihilations.
 The case of combined slepton and squark
 coannihilations gives a lower mass limit of a wino DM as 1~TeV.
However, in spite of the appearance of the above limits
we must keep in mind that there can be an 
order of magnitude of uncertainty in the computation 
of the SI direct detection cross-section. This may potentially lower the above 
mass limits by 10 to 15\%. The indirect detection data such as that from 
the IceCube for the muon flux do not put any additional constraint 
than whatever is given by the relic density and the SI direct detection
cross section data in combination. 
Regarding future experiments, XENON1T would be able to probe 
only the higgsino LSP scenario with both kinds of sfermion coannihilations.
Finally, with relevant bounds from ATLAS and CMS being satisfied,
  we pointed out that for LSP and $\chonepm$ either
  being a higgsino or a 
  wino dominated in nature there is hardly any collider bound to
  worry about while considering the compressed pMSSM scenario
  where the sfermion masses would be suitable for DM coannihilations.

{\bf Acknowledgments}
UC is thankful to receive the hospitality from CERN Theory division
where a major part of this work was completed. MC would like to thank
TRR33 "The Dark Universe" project for financial support.


\newpage{\pagestyle{empty}\cleardoublepage}


\end{document}